# High-Fidelity Transcranial Ultrasound Multi-focal Stimulation via Physics-Aware Hologram Technique

Moon Hwan Lee[1,4], Mohd. Afzal Khan[1], Akm Ashiquzzaman[2], Eunbin Lee[2],
Jonghun Lee[1], Euiheon Chung[2*], Hyuk-Sang Kwon[2*], Jae Youn Hwang[1,3*]
[1]Department of Electrical Engineering and Computer Science, DGIST, Daegu, 42988, South Korea.
[2]Department of Biomedical Science and Engineering, GIST, Gwangju, 61005, South Korea.
[3]Biomedical Engineering Program, DGIST, Daegu, 42988, South Korea.
[4]Present address: Max Planck Institute for Medical Research, Heidelberg, 69120, Germany.
[*]Corresponding author. Email: ogong50@gist.ac.kr, hyuksang@gist.ac.kr, jyhwang@dgist.ac.kr

***Introduction:*** **Transcranial ultrasound stimulation (TUS) is an emerging non-invasive neuromodulation modality that offers deep brain access with high spatial precision. However, its broader application is limited by the difficulty of reliably generating complex transcranial acoustic fields, particularly for multi-target stimulation through the skull. These limitations can lead to focal distortion, off-target exposure, and reduced reliability of neuromodulation outcomes.** ***Materials and Methods:*** **Here, we introduce a physics-aware hologram technique for precise transcranial ultrasound neuromodulation. Unlike conventional approaches that rely on simplified phase-based approximations, the proposed method generates fabrication-ready holographic implementations while preserving consistency between numerical field synthesis and physical acoustic realization. This enables accurate formation of single-, dual-, and tri-focal stimulation patterns under transcranial conditions and improves localized targeting for multi-region neuromodulation. We validated the proposed technique through in silico simulations, ex vivo acoustic measurements through skulls, and in vivo experiments.** ***Results:*** **Compared with state-of-the-art methods, the proposed approach achieved improved focal reconstruction, enhanced energy confinement at intended targets, and reduced off-target acoustic leakage. To further assess functional relevance, we applied this approach in a neuropathic pain mouse model using simultaneous bilateral stimulation of the thalamic nuclei. The stimulation induced measurable changes in neuronal activity, reflected by reduced c-Fos expression, together with preliminary improvements in pain-related behavioral responses. These findings support the capability of the proposed technique to enable spatially localized and reproducible neuromodulation in vivo.** ***Conclusion:*** **Collectively, this work provides a practical proof-of-concept strategy for achieving high-precision, multi-target transcranial neuromodulation and supports further investigation of TOAH for neuroscience research and future therapeutic applications.**

# Introduction

Neuromodulation is transforming the treatment of neurological disorders by enabling targeted modulation of brain activity. This rapidly advancing field encompasses both implantable and non-invasive techniques aimed at treating a wide range of neurological conditions [1]. Among these, deep brain stimulation (DBS) is one of the most established approaches, providing therapeutic benefits through direct electrical stimulation of specific brain regions [2]. However, DBS requires invasive surgical implantation, which carries risks such as hemorrhage and infection.

To overcome these limitations, non-invasive neuromodulation techniques—including transcranial magnetic stimulation (tMS), transcranial current stimulation (tCS), and transcranial focused ultrasound (tFUS)—have been actively explored. Among these modalities, **transcranial focused ultrasound (tFUS)** is particularly promising due to its ability to reach deep brain targets with millimeter-scale spatial resolution, making it a compelling non-invasive alternative to DBS [3]. Although the underlying neuromodulatory mechanisms remain under active investigation, involving both thermal and mechanical pathways [4], [5], numerous in vitro and in vivo studies have demonstrated that ultrasound can modulate neural activity in a parameter-dependent manner [6], [7]. In addition to neuromodulation, tFUS has been investigated for a range of therapeutic and diagnostic applications, including targeted drug delivery and ultrasound-mediated blood–brain barrier opening [8], [9]. Despite these advances, several challenges remain for broader research and clinical translation. In particular, device-level limitations—such as skull-induced acoustic field distortion and the limited ability to achieve flexible and accurate spatial targeting—continue to restrict the predictability, reliability, and spatial degrees of freedom in transcranial ultrasound stimulation.

Recently, acoustic holography has emerged as a promising approach for enabling precise and flexible transcranial neuromodulation. By encoding complex wavefronts onto phase plates or hologram lenses, this approach enables the reconstruction of desired acoustic fields when coupled to a single-element ultrasound transducer [10]. First, it allows compensation for skull-induced aberrations by incorporating subject-specific acoustic properties and geometries of the skull into the lens design. This capability has been demonstrated across multiple models, including small animals, non-human primates, and humans [11], [12], [13]. Second, acoustic holography enables multi-focal stimulation by generating multiple target foci simultaneously [14], [15], [16], [17], providing a compact and accessible alternative to multi-channel systems. These capabilities facilitate spatially localized, multi-region neuromodulation that is otherwise difficult to achieve with conventional single-focus tFUS. Recent studies have demonstrated applications such as multi-region blood–brain barrier opening, neuromodulation of distributed circuits, and reduced activation thresholds through spatially encoded stimulation patterns [11], [18], [19].

Despite these advances, a key limitation remains in current acoustic holography, limiting its ability to achieve reliable and precise transcranial neuromodulation. Acoustic fields are typically designed using simplified representations (two-dimensional phase delay maps), while their physical realization involves complex three-dimensional hologram structures [10], [13]. This mismatch leads to discrepancies between the intended and delivered acoustic fields, particularly under transcranial and multi-focal conditions [20], [21]. As a result, targeting accuracy can degrade and stimulation outcomes become less predictable, limiting the reliability of neuromodulation. Such inconsistencies may lead to unintended off-target stimulation, complicating the interpretation of experimental results and raising safety concerns in transcranial applications [12].

Here, we address this limitation by introducing a physics-aware design method for precise and reliable transcranial neuromodulation. Rather than relying on the simplified representations, the proposed approach directly accounts for the physical propagation of acoustic waves during the design process, enabling more accurate and predictable spatial targeting through the skull. Specifically, we implement a thickness-only acoustic hologram (TOAH), in which the design parameters are jointly optimized together with the acoustic properties of the heterogeneous medium, including the subject's skull and the hologram lens itself. By incorporating physical propagation effects and geometric constraints into a unified optimization process, the resulting acoustic field remains more consistent with the intended neural targets.

Using multi-focal stimulation in a murine model, we demonstrate that this technique enables precise and spatially localized transcranial ultrasound stimulation, supported by comprehensive in silico and ex vivo validation. In vivo experiments further show that the designed acoustic fields can be reliably delivered to bilateral thalamic targets, inducing measurable behavioral and neuronal modulation. Collectively, this work introduces a practical technique for high-precision, multi-target neuromodulation through the intact skull, addressing a key challenge in non-invasive brain stimulation.

# Materials and Methods

## End-to-end physics-aware hologram lens optimization

Voxel-wise binary representations of the lens volume are intrinsically discontinuous (0: void, 1: solid), making them incompatible with automatic differentiation. To address this, we introduce a differentiable relaxation, termed *Differentiable Hologram Lens Approximation (DHLA)*(Fig. 1A), which maps a low-dimensional parameterization to a three-dimensional hologram lens. Specifically, DHLA defines a smooth mapping from a two-dimensional parameter map $\theta \in \mathbb{R}^{N_x \times N_y}$ to a quasi-binary volume $V' \in [0,1]^{N_x \times N_y \times N_v}$ while preserving gradient flow.

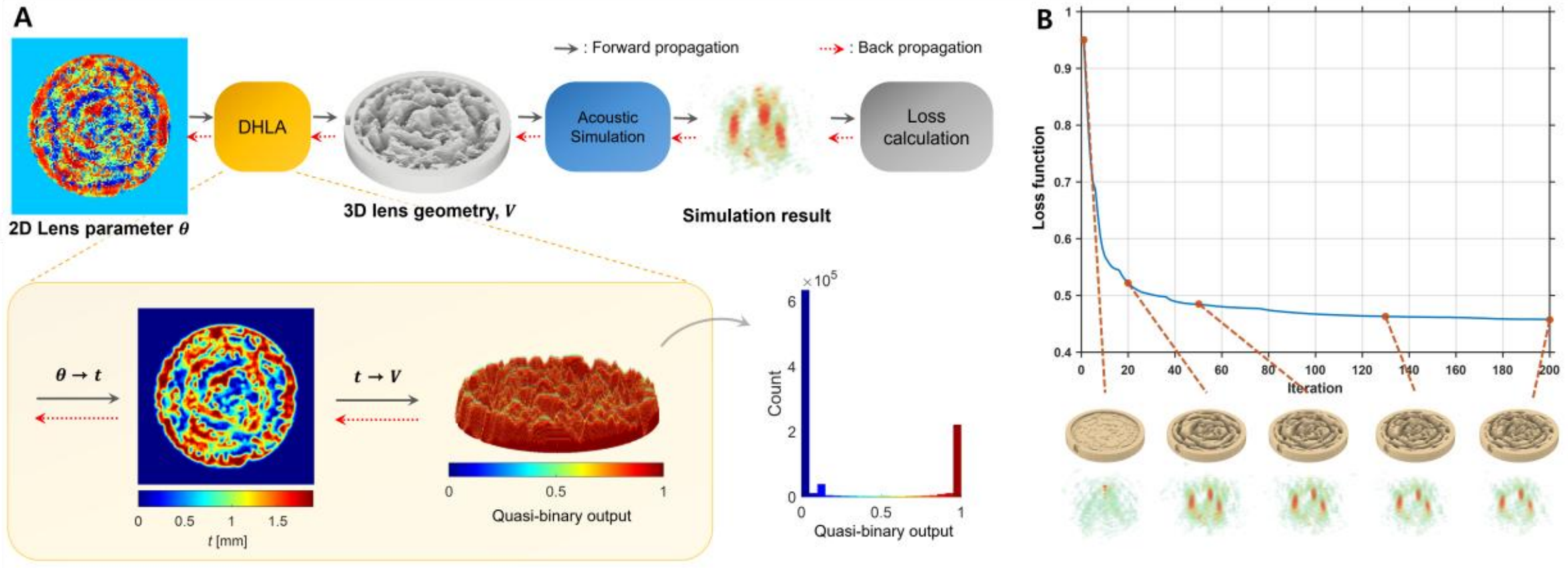

*Figure 1:* ***Differentiable hologram lens approximation and end-to-end physics-aware acoustic hologram optimization.*** *(**A**) End-to-end differentiable optimization pipeline for TOAH design. A continuous lens parameter map $\theta$ is mapped to a three-dimensional lens geometry via the DHLA, followed by acoustic wave simulation and loss calculation. The histogram illustrates the resulting quasi-binary distribution of the optimized lens. (**B**) Optimization trajectory of the proposed TOAH. The loss function decreases over iterations as the hologram geometry evolves to better reproduce the target acoustic field. Representative snapshots of the lens geometry and corresponding simulated acoustic fields illustrate the progressive improvement in reconstruction during optimization.*

The mapping $\Psi: \theta \rightarrow V'$ consists of three steps. First, the parameter map is converted to a continuous thickness field using a scaled sigmoid function,

$$t(i,j) = \sigma(\alpha \cdot \theta(i,j))\,(V_{\max} - V_{\min}) + V_{\min},$$

where $\sigma(\cdot)$ is the sigmoid function, $V_{\min}$ and $V_{\max}$ define the thickness range, and $\alpha$ controls smoothness.

Second, the thickness field is regularized via Gaussian smoothing to suppress high-frequency artifacts,

$$t_{\text{smooth}}(i,j) = (G * t)(i,j),$$

where $G$ is a Gaussian kernel and $*$ denotes convolution.

Finally, the smoothed thickness field is lifted to a three-dimensional quasi-binary volume using differentiable voxelization,

$$V'(i,j,k) = \sigma(\beta(k - t_{\text{smooth}}(i,j))), \qquad k = 1, \dots, N_v,$$

where $\beta$ controls the sharpness of the material transition. During optimization, $\beta$ is progressively increased to encourage convergence toward binary states while maintaining gradient flow (see Section S2 for pseudocode).

Through this composite mapping, DHLA ensures that the optimized lens remains monolithic and fabrication-compatible. The differentiable formulation allows gradients of the objective $\mathcal{L}$ to propagate from the acoustic field to the design parameters,

$$\frac{\partial \mathcal{L}}{\partial \theta} = \frac{\partial \mathcal{L}}{\partial V'} \cdot \frac{\partial V'}{\partial \theta},$$

enabling end-to-end optimization of the physical lens geometry as shown in Fig. 1B.

## Differentiable acoustic wave simulation

We adopt a mixed-domain acoustic propagation method (MDM) that accounts for medium heterogeneity, reflections, and diffraction while remaining compatible with automatic differentiation [22], [23], [24]. Wave propagation is modeled using a spatial-marching scheme that alternates between spatial and spectral domains, enabling efficient simulation in three-dimensional heterogeneous media. The formulation remains differentiable with respect to acoustic properties and geometric boundaries, allowing gradients to propagate through the wave solver during optimization.

For computational efficiency, we employ a frequency-specific formulation tailored to single-frequency hologram design [25]. This frequency-specific MDM (FSMDM) reduces computational cost compared to broadband time-domain solvers while maintaining sufficient accuracy for steady-state acoustic fields. The solver is implemented within a TensorFlow-based automatic differentiation framework, enabling seamless integration with the TOAH optimization pipeline.

The optimized lens geometry produced by DHLA is embedded into the heterogeneous medium by locally modifying voxel-wise acoustic properties, including sound speed, density, and attenuation (see Section S3). This approach explicitly represents the lens–medium interface during wave propagation while preserving gradient flow. As a result, the simulated acoustic field closely reflects the physical behavior of the fabricated hologram in a heterogeneous environment.

## Optimization Procedure

Let $\boldsymbol{U}$ denote the transducer parameters, including operating frequency and aperture geometry, and let $\boldsymbol{M}$ represent the heterogeneous acoustic medium incorporating the embedded lens geometry. The forward simulation yields a complex pressure field

$$P_{\text{sim}} = f(\boldsymbol{U}, \boldsymbol{M}) \ \in \ \mathbb{C}^{N_x \times N_y \times N_z},$$

where $f(\cdot)$ is the differentiable propagation operator described above. Given a target pressure distribution $P_{\text{target}}$, the hologram design problem is formulated as

$$V^* = \underset{V'}{\operatorname{argmin}} \ \mathcal{L}\big(P_{\text{sim}}, P_{\text{target}}\big),$$

where $V'$ is the quasi-binary lens volume generated by DHLA and $\mathcal{L}(\cdot)$ measures the discrepancy between simulated and target fields. In this work, we optimize the amplitude of the pressure field for transcranial stimulation, although the proposed method supports both amplitude and phase optimization.

The loss function incorporates multiple design objectives: (i) a field reconstruction term enforcing agreement with the target amplitude distribution, (ii) an energy efficiency term promoting energy concentration within focal regions, and (iii) a focal balance term ensuring uniform energy delivery across multiple targets. Detailed formulations are provided in Section S3.

Owing to the differentiable DHLA and wave propagation model, gradients with respect to the design parameters $\theta$ are obtained via the chain rule, enabling iterative updates using standard gradient-based optimizers (Algorithm S2, Section S3). This end-to-end optimization explicitly accounts for full-wave interactions at the lens–medium interface, including refraction and impedance-induced aberrations, which are neglected in conventional phase-only approaches.

## Hologram design conditions

The rodent skull geometry was obtained from a high-resolution micro-computed tomography (micro-CT) dataset [26]. Voxel-wise Hounsfield Unit (HU) values were converted into acoustic property maps, including sound speed, density, and attenuation coefficient, using experimentally calibrated relationships [27]. These spatially varying properties were embedded into the three-dimensional simulation domain to model wave propagation through the heterogeneous skull.

The simulation domain spanned $16\ \mathrm{mm} \times 16\ \mathrm{mm} \times 12\ \mathrm{mm}$ along the lateral ($x$), elevational ($y$), and axial ($z$) directions, respectively, and was discretized at six points per wavelength in water ($c = 1500\ \mathrm{m/s}$), corresponding to a spatial resolution of $125\ \mu\mathrm{m}$ at $2\ \mathrm{MHz}$. A single-element ultrasound transducer with a $13\ \mathrm{mm}$ aperture was positioned at the top of the domain and oriented along the axial direction.

The optimized thickness-only acoustic hologram was placed directly at the transducer aperture, with an acoustic coupling cone inserted between the hologram and the skull to ensure proper coupling. The acoustic properties of both the hologram and the coupling cone were assigned to match those of the 3D printing material (Clear resin, Formlabs, MA, USA; see Table S1) [28].

The skull position and target locations were defined using stereotaxic coordinates. The bregma was positioned $5.25\ \mathrm{mm}$ below and $2.25\ \mathrm{mm}$ anterior to the transducer center, ensuring consistent alignment across simulations and experiments. Target coordinates are summarized in Table S2.

Hologram thickness constraints were determined based on phase modulation induced by sound-speed mismatch between the lens material and water. To achieve a full $2\pi$ phase range at $2\ \mathrm{MHz}$, the maximum thickness was set to $1.78\ \mathrm{mm}$ (Eq. S5), while a minimum thickness of $250\ \mu\mathrm{m}$ was imposed for mechanical stability, yielding a manufacturable range of $250\ \mu\mathrm{m}$ to $2.03\ \mathrm{mm}$.

## In silico acoustic field simulations

Numerical validation was performed using the k-Wave MATLAB toolbox [29], a well-established toolbox for full-wave acoustic simulation in heterogeneous media that has been widely used for transcranial ultrasound treatment planning [30]. All simulation parameters, including transducer geometry, operating frequency, skull acoustic properties, and hologram geometry, were matched to those used during hologram optimization. The final acoustic medium was exported as a MAT file after the optimization in Python and imported into MATLAB to ensure a consistent simulation environment. The transducer was driven by a continuous-wave excitation at $2\ \mathrm{MHz}$ to generate steady-state fields.

To improve lateral resolution, the simulation grid was upsampled by a factor of two along the $x$ and $y$ directions, resulting in 12 points per wavelength (PPW), while maintaining 6 PPW in the axial direction. Perfectly matched layers were applied at the domain boundaries to suppress artificial reflections. The resulting pressure fields were used for quantitative evaluation.

To assess numerical–physical mismatch, acoustic fields were simulated in both the optimization and fabrication domains for each method. For phase-based approaches (e.g., Diff-PAT and time reversal), optimization-domain fields were obtained by applying phase delays, whereas fabrication-domain fields were computed using corresponding thickness-modulated lenses. In contrast, for physics-based approaches (PhyAH and TOAH), hologram structures were modeled consistently in both domains.

To account for finite 3D printing resolution, a spatial smoothing filter was applied to all lens geometries in the fabrication domain [10], [20]. This procedure reflects realistic fabrication processes and enables evaluation of reconstruction errors induced by manufacturing-related geometric deviations.

## Evaluation metrics for in silico study

To quantitatively evaluate the performance of the TOAH, we assessed acoustic field reconstruction accuracy and energy efficiency using the following metrics. All evaluations were conducted under identical geometric and acoustic conditions to ensure fair comparison across hologram design methods.

### *(1) Cross-domain fidelity:*

Cross-domain fidelity was defined as the similarity between acoustic fields in the optimization and fabrication domains and was quantified using the peak signal-to-noise ratio (PSNR) [31], with the optimization-domain field as the reference. Higher PSNR indicates better preservation of the designed acoustic field after fabrication-domain conversion.

*(2) Focal characteristics:*

Focusing performance was characterized using the pressure amplitude field in the fabrication domain. The field was binarized using a $-6$ dB threshold relative to the peak pressure, followed by region growing from the target locations to extract the corresponding focal volumes. From these regions, axial and lateral full width at half maximum (FWHM) and focal volume were computed.

*(3) Energy confinement:*

Energy confinement was quantified using an energy leakage ratio, defined as the ratio of mean pressure amplitude outside to that inside the segmented $-6$ dB focal volume. Lower values indicate improved energy concentration within the target region. For multi-focal cases, intensity uniformity was additionally evaluated as the ratio between the minimum and maximum peak focal intensities ($P_{\text{max}}^2$) across targets, reflecting balanced energy delivery.

## Fabrication of holographic ultrasound module and acoustic holograms

The holographic ultrasound module consists of three components: a custom-built single-element transducer, a 3D-printed hologram lens, and a screw-cap acoustic coupling cone. The hologram lenses, transducer housing, and coupling cone were fabricated using a Form 3+ SLA printer (Formlabs, Somerville, MA, USA).

The transducer was fabricated following a modified rapid prototyping protocol [32]. A piezoelectric ceramic disk (PZT-4, Dongiltech, South Korea; 13 mm diameter, 2 MHz center frequency) served as the active element. Silver electrodes were deposited on both surfaces in a wraparound configuration, and a micro-coaxial cable was soldered with the ground and signal connected to the front and rear electrodes, respectively. The PZT disk was inserted into the housing and secured with UV epoxy.

To improve transmission efficiency across the impedance mismatch between the PZT (~33 MRayl) and the hologram lens material (Clear resin, 2.922 MRayl), an aluminum-oxide–epoxy matching layer (6.7 MRayl [33]) was applied to the front surface following established impedance-matching principles [34]. The layer thickness was defined based on PiezoCAD design specifications. An air-backed condition was created by sealing the rear of the housing, enhancing forward acoustic radiation (Fig. S7).

For module assembly, the hologram lens was immersed in water and gently flushed to remove trapped air bubbles. It was then inserted into the coupling cone, which incorporated a key–slot alignment feature to enforce a unique orientation consistent with the design. The coupling cone was filled with degassed ultrasound gel and attached to the transducer via a screw-cap mechanism. The gel remained within the cone during assembly due to its high viscosity.

## Experimental setup for ex-vivo acoustic field measurement

Ex vivo measurements were conducted to evaluate the reconstructed acoustic fields generated by the proposed thickness-only acoustic hologram (TOAH). Fig. S1 illustrates the experimental setup. A harvested mouse skull was mounted on a custom-designed fixture based on its micro-CT scan (Bruker, SkyScan 1176) to replicate the transducer–skull geometry used in simulations.

A single-element ultrasound transducer (2 MHz center frequency, 13 mm aperture) was driven by a function generator and a 50U1000 power amplifier (Amplifier Research, PA, USA). Acoustic pressure was measured using a needle hydrophone (0.2 mm diameter, Precision Acoustics, UK) mounted on a three-axis motorized stage (OSMS26, OptoSigma, USA). The acquired signals were digitized using a CSE1622 DAQ board (GaGe Applied, Canada).

Direct volumetric scanning within the skull cavity was not feasible due to spatial constraints and the risk of hydrophone–skull collision. In addition, full 3D scanning would require prohibitively long acquisition times. Instead, acoustic pressure was measured on a single plane beneath the skull, and the full three-dimensional field was reconstructed by backpropagating the measured complex pressure using the angular spectrum method (see Section S4).

This projection-based measurement approach enables recovery of volumetric acoustic fields while reducing experimental risk and acquisition time, and has been shown to agree well with direct volumetric measurements [35], [36].

## Animal preparation and neuropathic pain model

All experimental procedures were approved by the Institutional Animal Care and Use Committee (IACUC) of the Gwangju Institute of Science and Technology (Approval No. GIST-2024-028). A total of 12 male C57BL/6J mice (8 weeks old, 20–25 g) were used. Animals were randomly assigned to four groups (n = 3 per group): sham control (SC), SC+US, CCI, and CCI+US. Mice were housed under standard conditions with a 12 h light/dark cycle and provided ad libitum access to food and water. All behavioral assessments were conducted between 09:00 and 18:00. All procedures adhered to the AVMA Guidelines for the Euthanasia of Animals (2020) and were reported in accordance with the ARRIVE guidelines (https://arriveguidelines.org).

Neuropathic pain was induced using the chronic constriction injury (CCI) model of the sciatic nerve. Under anesthesia, the sciatic nerve was exposed at the mid-thigh level and loosely ligated with three 6-0 silk sutures spaced approximately 1 mm apart (Fig. S8). Sham groups underwent nerve exposure without ligation. Animals were allowed to recover before behavioral assessment.

Mechanical withdrawal thresholds were measured using the von Frey test on days $-1, 3, 5, 7, 9, 11,$ and 14 to track pain development (Fig. S9). Withdrawal thresholds on the contralateral paw were also assessed at the same time points (Fig. S9). No significant differences were observed on the contralateral side throughout the experimental period.

## In vivo experimental procedures

Animals were anesthetized via intraperitoneal injection of a Zoletil/Xylazine mixture (60/10 mg/kg). Anesthesia depth was confirmed using the tail pinch reflex. Neodex ophthalmic ointment (Hanlim Pharmaceuticals, South Korea) was applied to prevent corneal drying, and body temperature was maintained using a heating pad. Scalp hair was removed using an electric shaver followed by depilatory cream, taking care to avoid skin damage.

For ultrasound stimulation, animals were positioned in a stereotaxic frame (RWD Ltd., China) under anesthesia. Note that any anesthesia-related effect is equal across groups. On the stereotaxic frame, the bregma and lambda were identified to establish a flat-skull position. A stereotaxic-guided ultrasound stimulation scheme [37], [38] was adopted and modified for TOAH-enabled multi-focal stimulation. A custom-designed pointer was used for transducer alignment, following the approach of [38], with its tip indicating the bregma as the reference point for both lens design and experimental procedures.

The assembled transducer module was placed on the scalp with ultrasound coupling gel and mounted onto the stereotaxic holder. A magnetic attachment system enabled rapid switching between the pointer and the transducer assembly. Care was taken to avoid air bubble entrapment at all acoustic interfaces.

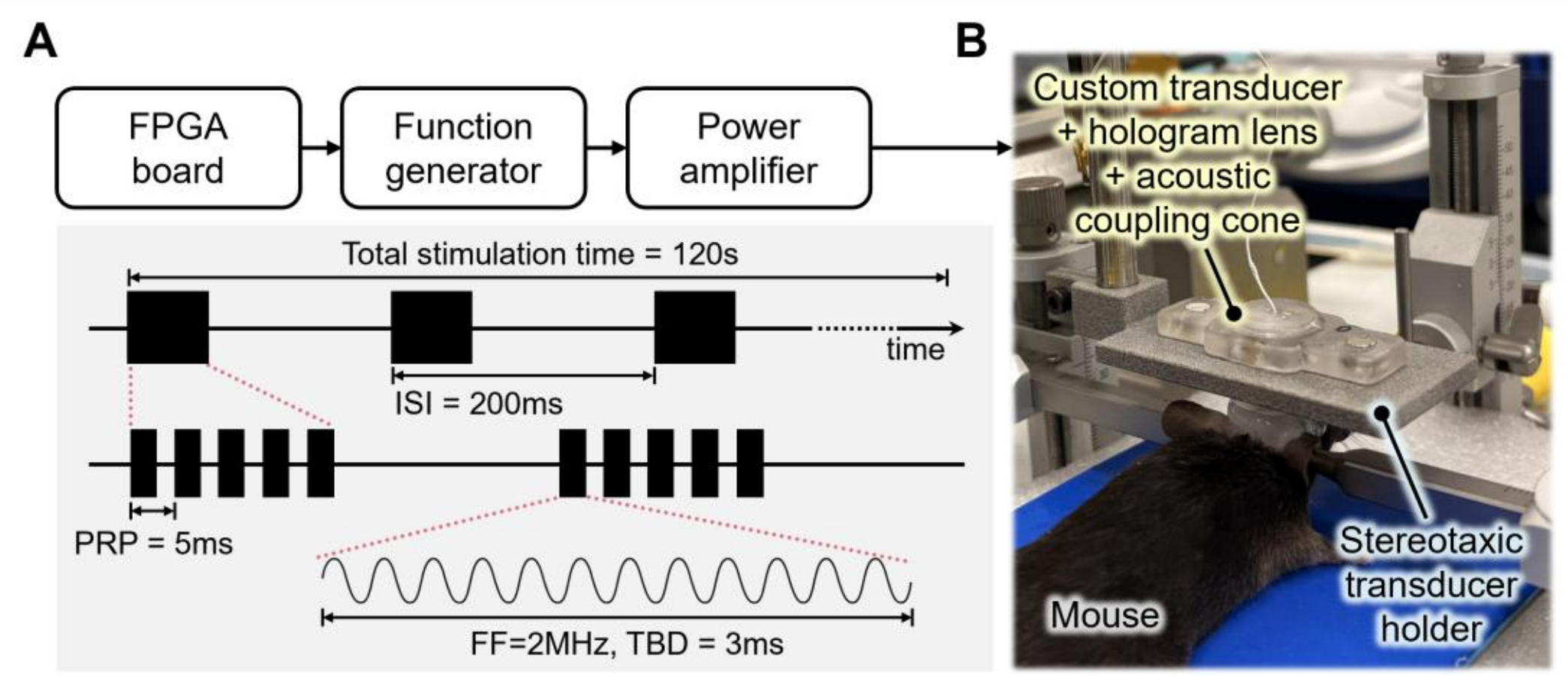

*Figure 2:* ***In vivo ultrasound stimulation protocol and experimental setup for TOAH-enabled multi-focal neuromodulation.*** *(**A**) Block diagram and temporal structure of the stimulation waveform. An FPGA-generated trigger gated a function generator to produce a 2MHz sinusoidal signal, which was amplified and delivered to the transducer. Stimulation was applied for 120s using repeated tone bursts (total burst duration (TBD) = 3ms, pulse repetition period (PRP) = 5ms) with an inter-stimulation interval (ISI) of 200 ms. (**B**) Photograph of the stereotaxic-guided stimulation setup. The holographic ultrasound module was positioned above the mouse head.*

Ultrasound stimulation was delivered for 2 min (Fig. 2) using parameters adapted from [39]. An FPGA board (Nexys 7, Digilent, WA, USA) generated trigger pulses to gate a function generator (Tektronix, OR, USA), producing a sinusoidal waveform that was amplified by a power amplifier (50U1000, Amplifier Research, PA, USA) to drive the transducer.

The peak negative pressures at the dual foci were calibrated to 0.75 MPa (left) and 0.73 MPa (right) using ex vivo measurements, corresponding to mechanical indices of 0.53 and 0.52, respectively. The spatial-peak temporal-average intensities ($I_{\mathrm{SPTA}}$) were estimated to be 1.13 W/cm$^2$ and 1.07 W/cm$^2$. To target different brain regions, only the hologram lens can be replaced while all other procedures remained unchanged.

## Behavioral assessment

Mechanical sensitivity was assessed using calibrated von Frey filaments. Mice were placed individually in transparent Plexiglas chambers on an elevated wire mesh platform and allowed to acclimate for at least 1 h. Mechanical stimuli were applied perpendicularly to the plantar surface of both the ipsilateral and contralateral hind paws using filaments of increasing force. Each filament was applied until slight bending and maintained for 4–5 s. A positive response was defined as rapid paw withdrawal, licking, or shaking. The paw withdrawal threshold (PWT) was determined using the ascending method and expressed in grams. All behavioral assessments were performed by an experimenter blinded to treatment conditions.

## Immunofluorescence staining

Mice were anesthetized and transcardially perfused with phosphate-buffered saline (PBS), followed by 4% paraformaldehyde (PFA). Brains were dissected and post-fixed in 4% PFA at 4°C overnight, then cryoprotected in 30% sucrose (Sigma-Aldrich, S8501) in PBS for 48 h.

Brains were sectioned at 25 $\mu$m thickness using a cryostat (Leica CM1850) after embedding in Tissue-Tek O.C.T. compound (Sakura, 4583). Sections were blocked in 5% goat serum with 0.3% Triton X-100 in PBS for 3 h at room temperature, followed by incubation with primary antibodies overnight at 4°C and secondary antibodies for 2 h at room temperature.

Primary antibodies included mouse anti-NeuN (1:500, Abcam) and rabbit anti-c-Fos (1:500, Thermo Fisher). Sections were mounted using ProLong Gold antifade reagent (Invitrogen) and imaged with a confocal microscope (Olympus FV1000).

## In silico evaluation in human skull models

For the human skull evaluation, a single-element ultrasound transducer with a $90\ \mathrm{mm}$ aperture diameter operating at a center frequency of $500\ \mathrm{kHz}$ or $1\ \mathrm{MHz}$ was modeled. The aperture diameter was selected to provide skull coverage comparable to the relative coverage used in the rodent experiments. Dual-foci and tri-foci target configurations, analogous to those used in the rodent experiments, were evaluated.

To assess the robustness of the proposed framework to inter-subject variation in human skull geometry, five human skull CT datasets were collected: four datasets (skulls A–D) from [40] and one dataset (skull E) from [41]. The transducer was positioned above the outer calvarial surface, and acoustic waves were transmitted through the calvaria in the superior-to-inferior direction (Fig. S11), following a geometry similar to that used in a previous transcranial simulation study [40].

As in the rodent-skull model, voxel-wise HU values were converted into acoustic property maps for hologram optimization and full-wave acoustic simulations. The computational domain spanned $100 \text{ mm} \times 100 \text{ mm} \times 110 \text{ mm}$ along the lateral, elevational, and axial directions, respectively. The simulation domain was discretized at six points per wavelength (PPW) for the $500 \text{ kHz}$ and three PPW for the $1 \text{ MHz}$ in water ($c = 1500 \text{ m/s}$), while maintaining comparable computational memory requirements.

A center frequency of $500 \text{ kHz}$ was selected because previous human transcranial ultrasound studies commonly employed similar low frequencies to mitigate skull-induced attenuation and improve acoustic penetration through the human skull [9], [13]. A center frequency of $1 \text{ MHz}$ was additionally investigated as a more challenging simulation condition to assess whether TOAH could maintain accurate multi-focal reconstruction under increased skull-induced phase aberration while potentially enabling higher spatial resolution. Although 1 MHz is less commonly used for human transcranial ultrasound applications because of increased skull attenuation and aberration, this condition provides a stringent test of the framework's ability to compensate for complex transcranial distortions.

## Statistical Analysis

For the in vivo animal experiment result, data are presented as mean $\pm$ standard deviation, with individual animals shown where applicable. For four-group comparisons, statistical analysis was performed using two-way ANOVA with disease status (sham or CCI) and ultrasound treatment (without US or US) as factors, followed by Sidak's multiple-comparisons test.

For comparisons between TOAH and TR in the human-skull simulations, linear mixed-effects models were used, with hologram design method, ultrasound frequency, and focal configuration (number of foci) specified as fixed effects and skull sample specified as a random effect. Statistical significance was set at $P < 0.05$.

# Results

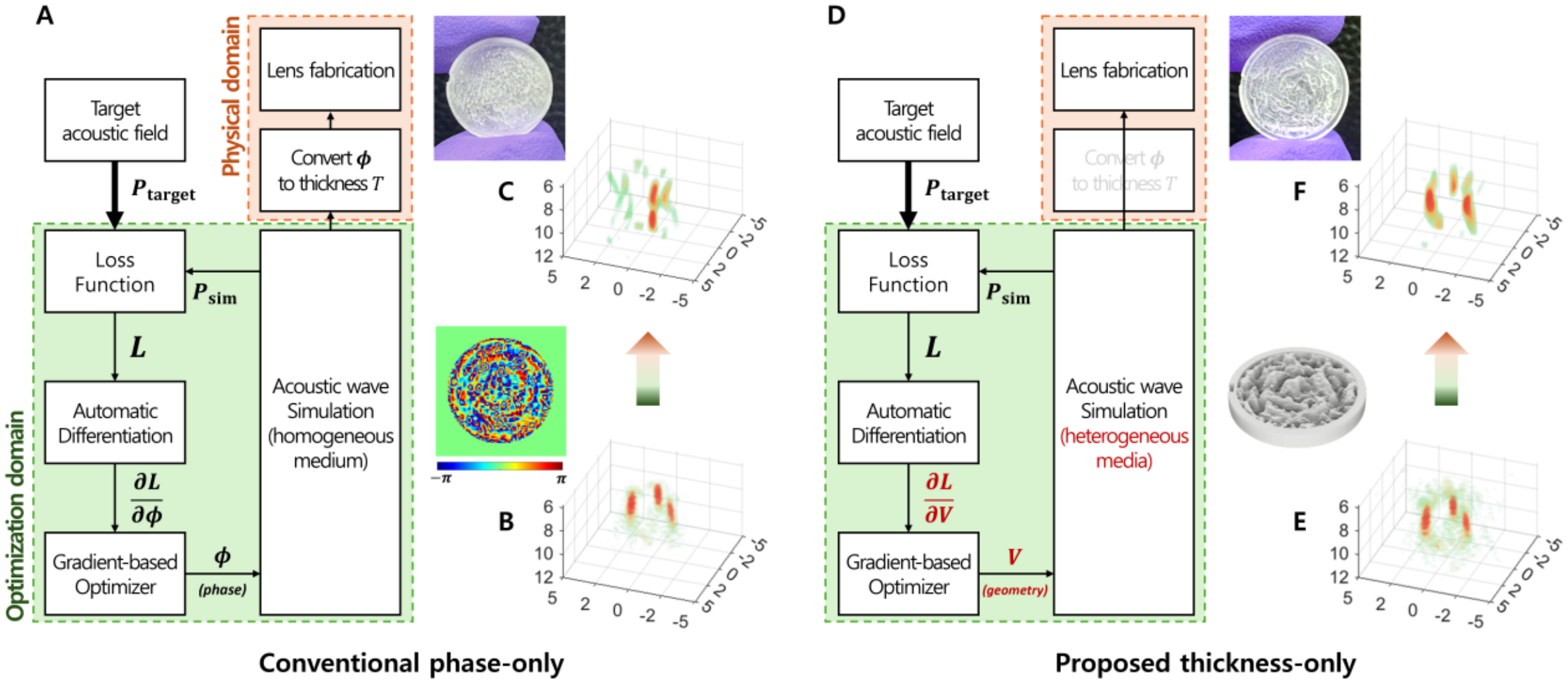

*Figure 3:* ***Conventional simplified assumption-based phase-only versus proposed physics-aware thickness-only acoustic hologram****. (**A, D**) Pipelines of the conventional phase-only and the proposed thickness-only acoustic hologram. (**B, E**) Optimized holograms and simulated acoustic fields in the phase-only and thickness-only hologram methods, respectively. (**C, F**) Implemented holograms and measured acoustic fields in physical domain where the holograms are fabricated as lenses.*

Fig. 3 illustrates the fundamental difference between conventional acoustic hologram design and the proposed approach. Conventional methods derive two-dimensional phase-delay maps, $\phi$ (Fig. 3B, left), to reconstruct target acoustic fields, but implement them as thickness-varying volumetric lenses (Fig. 3C, left). This mismatch neglects wave distortions introduced by the volumetric lens geometry, leading to discrepancies between the designed and physically realized acoustic fields (Fig. 3B, right vs. C, right), particularly under transcranial or multi-focal conditions.

In contrast, the proposed method directly represents the hologram as a thickness-varying volumetric lens $V$, consistent with its physical implementation (Fig. 3E, left; F, left). By incorporating these physical effects during the design process, TOAH enables accurate reconstruction of the target acoustic field in the physical domain, as confirmed by experimental measurements (Fig. 3E, right; F, right). Detailed descriptions of the proposed optimization framework are provided in the Materials and Methods section.

To demonstrate the high-fidelity transcranial multi-focal stimulation capability of the proposed TOAH, we conducted comprehensive evaluations, including in silico experiments to efficiently evaluate multiple cases, ex vivo experiments to validate practical transcranial ultrasound focusing, and in vivo experiments to demonstrate its potential as a tool for precise and reliable transcranial ultrasound multi-focal stimulation.

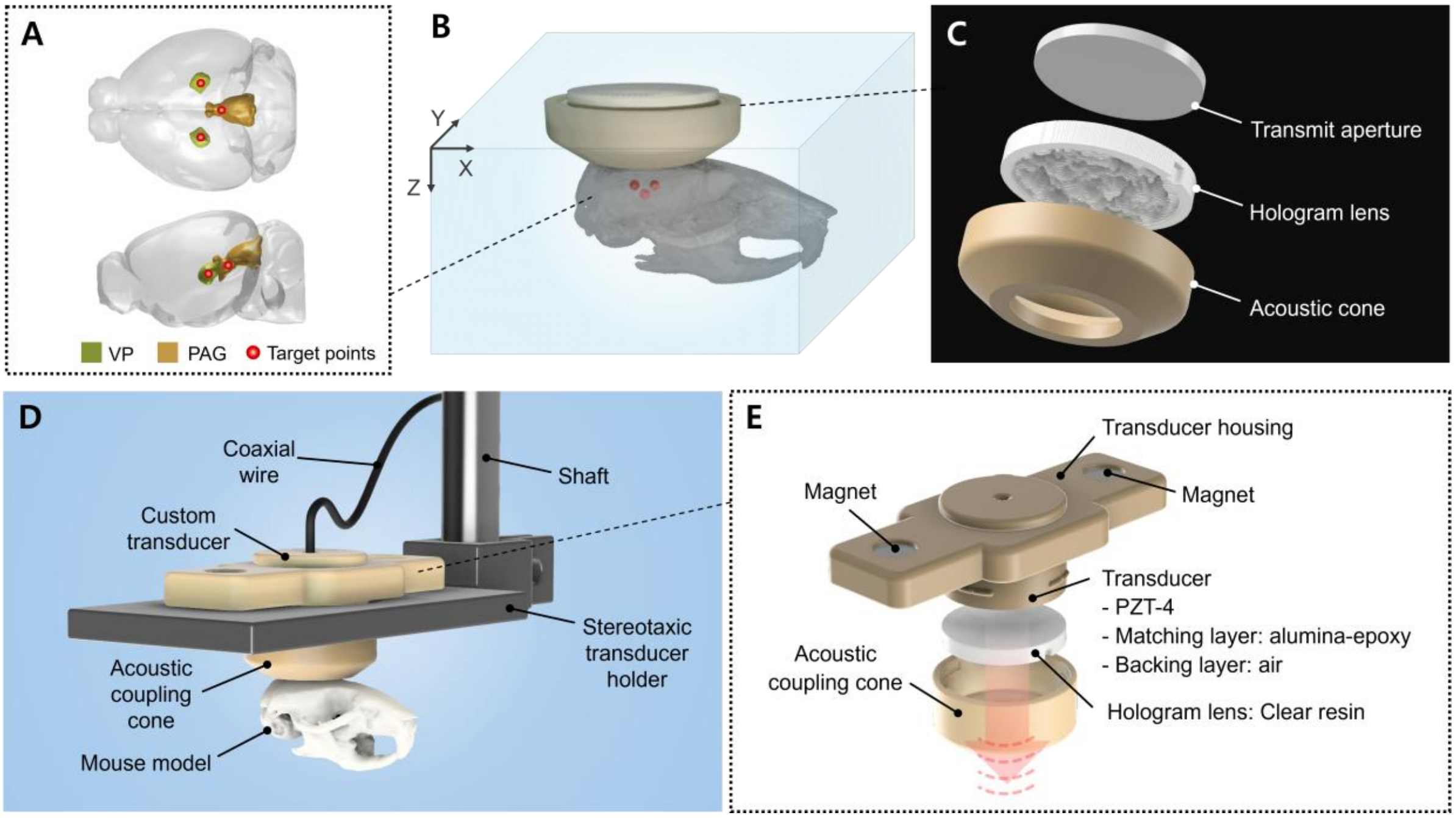

*Figure 4:* ***Transcranial ultrasound stimulation setup using the proposed TOAH.*** *(**A**) Target regions for multi-focal stimulation in the mouse brain, including the ventral posterior nuclei (VP) and periaqueductal gray (PAG). (**B**) Numerical simulation setup for transcranial ultrasonic focusing through a rodent skull model. Red circles indicate the target focal points. (**C**) Exploded view of the holographic ultrasound module consisting of a transmit aperture, the hologram lens, and an acoustic coupling cone. (**D**) Experimental setup for transcranial stimulation using a stereotaxic transducer holder and a custom ultrasound transducer aligned with the rodent model. (**E**) Structural design of the transducer assembly showing the PZT-4 disk, matching layer (alumina–epoxy), air backing layer, and the hologram lens fabricated via 3D printing.*

Fig.4 illustrates the experimental setup used in the evaluations. Fig.4A shows the deep brain target regions in a 3D mouse brain, including the ventral posterior nuclei (VP) of the thalamus and the periaqueductal gray (PAG). Fig.4B illustrates a case of simultaneous targeting of three regions through the mouse skull using the holographic ultrasound module (Fig.4C). The holographic ultrasound module comprises a planar transducer aperture (13mm diameter, 2MHz), a designed hologram lens, and an acoustic coupling cone. Notably, the acoustic coupling cone, as well as the skull, is explicitly incorporated into the lens design, highlighting the ability of the proposed technique to account for the actual experimental environment.

## Transcranial multi-focal focusing via thickness-only acoustic hologram (in silico)

Here, through in silico experiments, we evaluate the precise and reliable transcranial focusing capability of the proposed TOAH technqiue for spatially localized ultrasound stimulation in rodents and compare it with three state-of-the-art (SoTA) methods (Table. 1). Specifically, we systematically assess performance by progressively increasing target field complexity: *single-focus* targeting of the PAG, *dual-foci* targeting of the bilateral VP nuclei, and *tri-foci* targeting of the PAG and bilateral VP nuclei. As four hologram design methods were compared, a total of 12 lenses were evaluated.

**Table 1: Comparison of conventional acoustic hologram design methods and the proposed TOAH.** ***(TR: Time-reversal, PhyAH: Physics-based acoustic hologram)***

| Criteria | Diff-PAT[9] | TR[10] | PhyAH[11] | TOAH (This work) |
|---|---|---|---|---|
| Heterogeneous media | ✗ | ✓ | ✓ | ✓ |
| Gradient-based optimization | ✓ | ✗ | ✓ | ✓ |
| Optimization domain | Phase profile | Phase profile | Material composition | Thickness profile |
| Fabrication domain | Thickness-modulated plate | Thickness-modulated plate | Material-modulated plate | Thickness-modulated plate |
| Domain consistency | ✗ | ✗ | ✓ | ✓ |
| Accessibility | ✓ | ✓ | ✗ | ✓ |

The resulting acoustic fields are summarized in Fig. 5. Note that the optimization domain refers to simulations in which holograms are modeled in the same form as they are designed, as shown in Table 1 (i.e., Diff-PAT as phase-delay maps). In contrast, the fabrication domain refers to simulations in which holograms are modeled as they are physically implemented in real-world settings (i.e., Diff-PAT as thickness-modulated plates). While all designs reproduce the intended focal patterns in the optimization domain, significant degradation is observed when phase-only holograms are implemented as thickness-modulated lenses. The reconstructed fields exhibit elevated sidelobes and focal imbalance, with degradation becoming more pronounced as the number of target foci increases. This discrepancy stems from fabrication-induced wave distortions not accounted for in the simplified phase-domain optimization. In contrast, the proposed method preserves strong agreement between simulation and physical realization across all configurations.

This behavior is quantified in Fig. 5Ba, which reports the cross-domain field accuracy measured by the peak signal-to-noise ratio (PSNR) between acoustic fields in the optimization and fabrication domains. Phase-only approaches (Diff-PAT and TR) exhibit a substantial drop in cross-domain field accuracy, indicating that improvements achieved in the phase-only optimization domain do not reliably translate to the fabricated lens. In contrast, approaches that incorporate the physical lens geometry during optimization (PhyAH and TOAH) preserve strong agreement between optimized and fabricated acoustic fields. The isovolumes above $-6$ dB and $-3$ dB further demonstrate that the proposed TOAH accurately reconstructs the desired acoustic field, even as target field complexity increases (Fig. S3).

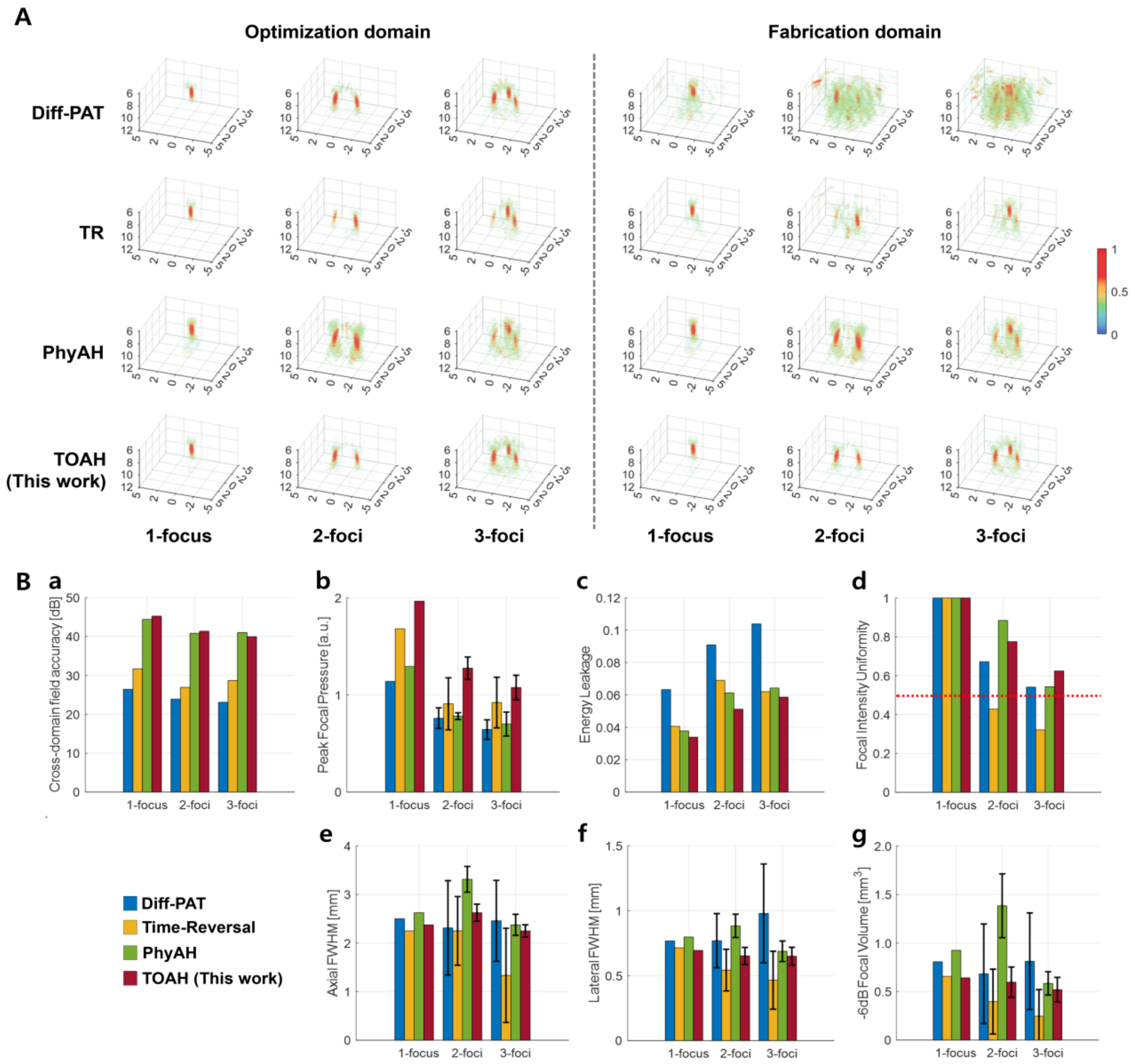

*Figure 5:* ***Consistency between optimization and fabrication domains.*** *(**A**) Reconstructed three-dimensional acoustic pressure fields in both domains for single-, dual-, and tri-focal configurations. (**B**) Quantitative performance metrics: (a) cross-domain field accuracy, (b) peak focal pressure, (c) energy leakage ratio, (d) focal intensity uniformity (red dashed line indicates the threshold for insufficient uniformity), (e,f) axial and lateral full width at half maximum (FWHM), and (g) -6 dB focal volume.*

We further evaluate focusing performance in the fabrication domain using several quantitative metrics (Fig. 5Bb–g), including peak focal pressure, energy leakage ratio, focal intensity uniformity, axial and lateral full width at half maximum (FWHM), and the $-6$ dB focal volume. Details of the metric calculations are provided in the Materials and Methods. Across all target field configurations, TOAH achieves efficient energy confinement within the target regions, maintaining high focal pressure and low energy leakage while preserving compact focal dimensions and balanced intensity across distributed targets.

Previous studies have shown that appropriate phase compensation can reduce thermal deposition in the highly absorptive skull, thereby mitigating safety concerns [42]. As an additional indicator of safe and effective transcranial focusing, we evaluate skull thermal deposition using the bioheat transfer equation (see Section S5 for details). Leveraging high-fidelity acoustic wave shaping, TOAH significantly reduces skull thermal deposition compared with conventional methods while maintaining equivalent acoustic pressure at the target foci (Fig. S4). These results demonstrate that TOAH preserves the designed wavefront, compensates for skull-induced distortions, and minimizes thermal deposition. In contrast, phase-only designs fail due to unmodeled distortions introduced by the volumetric lens.

Prior to ex vivo evaluation, we assess the robustness of holograms designed using the proposed technique under practical experimental variations, including driving pulse length, lens material properties (Table. S3), and fabrication-induced surface deviations. The results indicate that TOAH maintains stable focusing performance across realistic parameter ranges (Fig. S5, S6). It is noteworthy that sufficient pulse duration and accurate characterization of the acoustic impedance of the lens material are essential for preserving the designed acoustic field, whereas moderate variations in other parameters lead to only minor degradation and do not significantly affect performance.

## Ex vivo validation of thickness-only acoustic hologram

Next, we validate the proposed TOAH through ex vivo transcranial experiments using a harvested mouse skull. To ensure consistency between simulation and experiment, the in silico and ex vivo configurations are carefully matched, including the transducer geometry, hologram placement, and target configurations (Fig. 4D). Details of the experimental setup are provided in the Materials and Methods. We compare TOAH with Diff-PAT, a representative phase-based design approach. As both methods share the same thickness-modulated physical implementation and differ only in their optimization domains (Table 1), this comparison directly isolates the impact of TOAH's unified design–fabrication framework.

Fig. 6 presents experimentally measured acoustic fields for single-, dual-, and tri-focal configurations. Although both methods aim to reconstruct identical target fields, clear differences are observed in the measured acoustic patterns. Diff-PAT exhibits pronounced sidelobes, focal spreading, and imbalance across targets, particularly for multi-focal configurations. These distortions are consistent with the fabrication-domain degradation observed in the in-silico evaluations.

In contrast, TOAH consistently confines acoustic energy to the intended target regions across all configurations. The measured fields exhibit compact focal regions and balanced energy distribution, in close agreement with the corresponding in silico predictions. Notably, for the dual-foci configuration, volume asymmetries (left > right) predicted in the simulated fields are reproduced in the experimental measurements, further supporting the predictive fidelity of the physics-aware optimization framework.

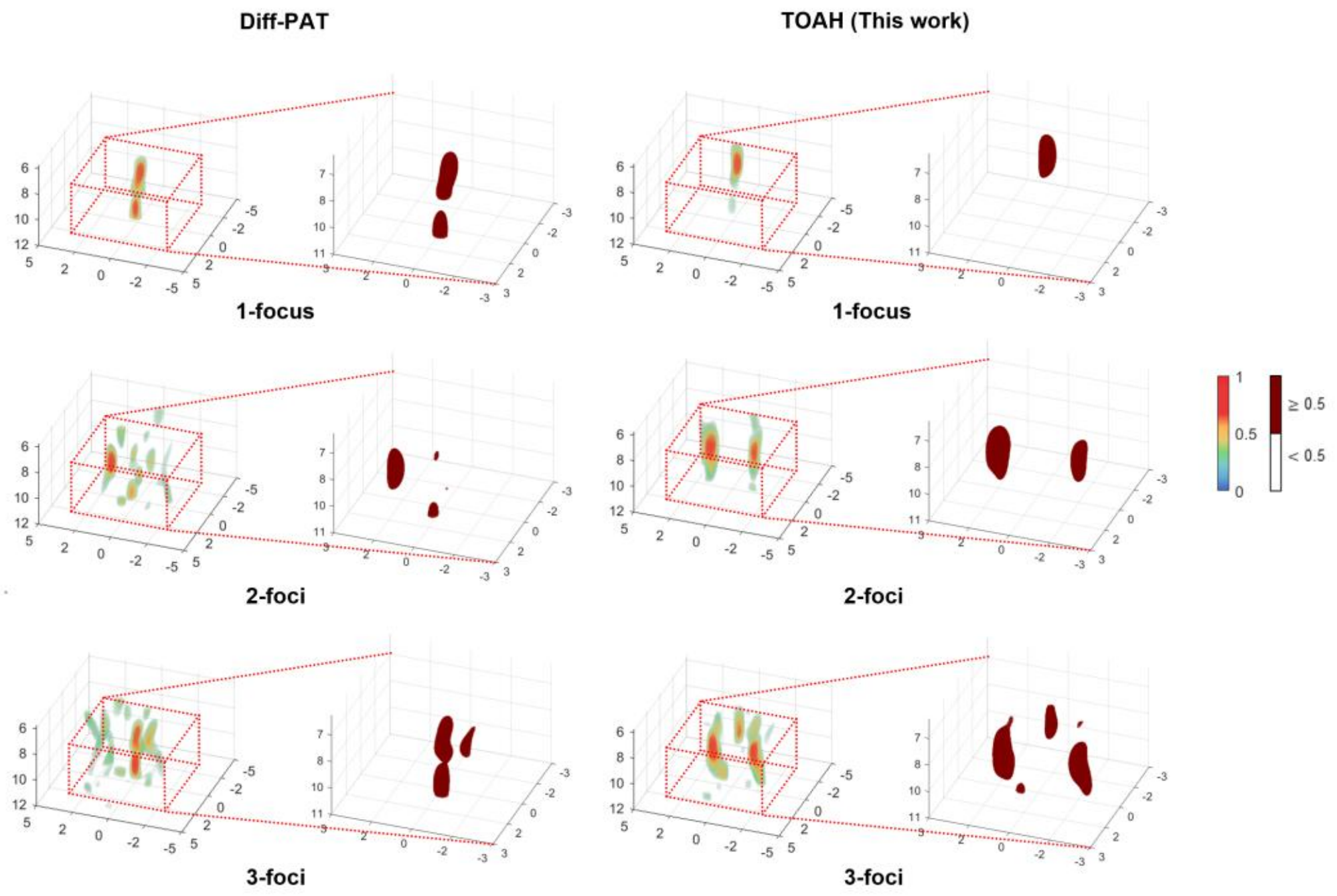

*Figure 6: **Ex vivo validation of transcranial acoustic field reconstruction**. Representative experimentally measured acoustic intensity fields for single-, dual-, and tri-focal configurations using Diff-PAT and the proposed TOAH. Left panels show the measured three-dimensional intensity distributions, and right panels show the corresponding –6 dB thresholded fields highlighting the reconstructed focal regions.*

These results demonstrate that the proposed hologram technique reliably bridges simulation and physical experiments, supporting more reliable acoustic field estimation and facilitating the interpretation of transcranial ultrasound multi-focal stimulation outcomes.

## In vivo transcranial neuromodulation with thickness-only acoustic hologram

Finally, we extended the proposed hologram technique to an in vivo demonstration to evaluate the practical feasibility of TOAH-enabled multi-focal transcranial ultrasound field delivery. Specifically, we stimulated the bilateral VP thalamic nuclei in a neuropathic pain mouse model induced by a chronic constriction injury (CCI) and analyzed both behavioral and neuronal responses. The VP nucleus is a well-established target for the treatment of neuropathic pain using DBS [43], [44]. A dual-foci TOAH lens (Fig. 7B), designed to reconstruct a spatially confined acoustic field (Fig. 7C) and validated through the preceding in silico and ex vivo experiments, was used to simultaneously target the bilateral VP nuclei (Fig. 7D). A total of 12 healthy male C57BL/6J mice (age, 8 weeks; weight, 20–25 g) were used in this study. Mice were randomly assigned to four groups (n = 3 per group): sham control (SC), SC+US, CCI, CCI+US.

The experimental timeline is illustrated in Fig. 7A. Briefly, ultrasound stimulation was delivered on days 8 and 11 following CCI surgery, and behavioral responses were evaluated using the von Frey test three days after each stimulation session (days 11 and 14). After the behavioral assessment on day 14, mice received a final ultrasound stimulation and were subsequently sacrificed for immunohistological analysis to quantify neuronal responses in the VP nuclei.

Mechanical withdrawal thresholds were significantly elevated in the CCI+US group compared with the CCI group on day 14 (Fig. 8B). Although not intended to establish therapeutic efficacy, this behavioral change supports that the TOAH-designed acoustic field can be reproducibly delivered in vivo and produce analgesia-associated neuromodulatory effects. In contrast, no significant differences were observed between the SC and SC+US groups under the present experimental conditions (Fig. S10).

Neuronal activity within the VP nuclei was further evaluated using the immediate early gene marker c-Fos (Fig. 8A). The density of $NeuN^+$ neurons did not differ significantly between the CCI and CCI+US groups (CCI: 14650 cells$/mm^2$, CCI+US: 15129 cells$/mm^2$, $P = 0.317$), indicating preserved neuronal integrity following ultrasound exposure (Fig. 8C). In contrast, c-Fos expression was significantly reduced in the CCI+US group relative to the CCI group (CCI: 15165.5 cells$/mm^2$, CCI+US: 13793.3 cells$/mm^2$, $P < 0.05$). Consistently, the proportion of active neurons—defined as the ratio of c-$Fos^+$ to $NeuN^+$ cells—was significantly lower following ultrasound stimulation (CCI: 51.7%, CCI+US: 43.9%, $P < 0.05$) (Fig. 8D). Between the SC and SC+US groups, no significant differences were observed in NeuN density, c-Fos expression, or the proportion of active neurons under the present experimental conditions.

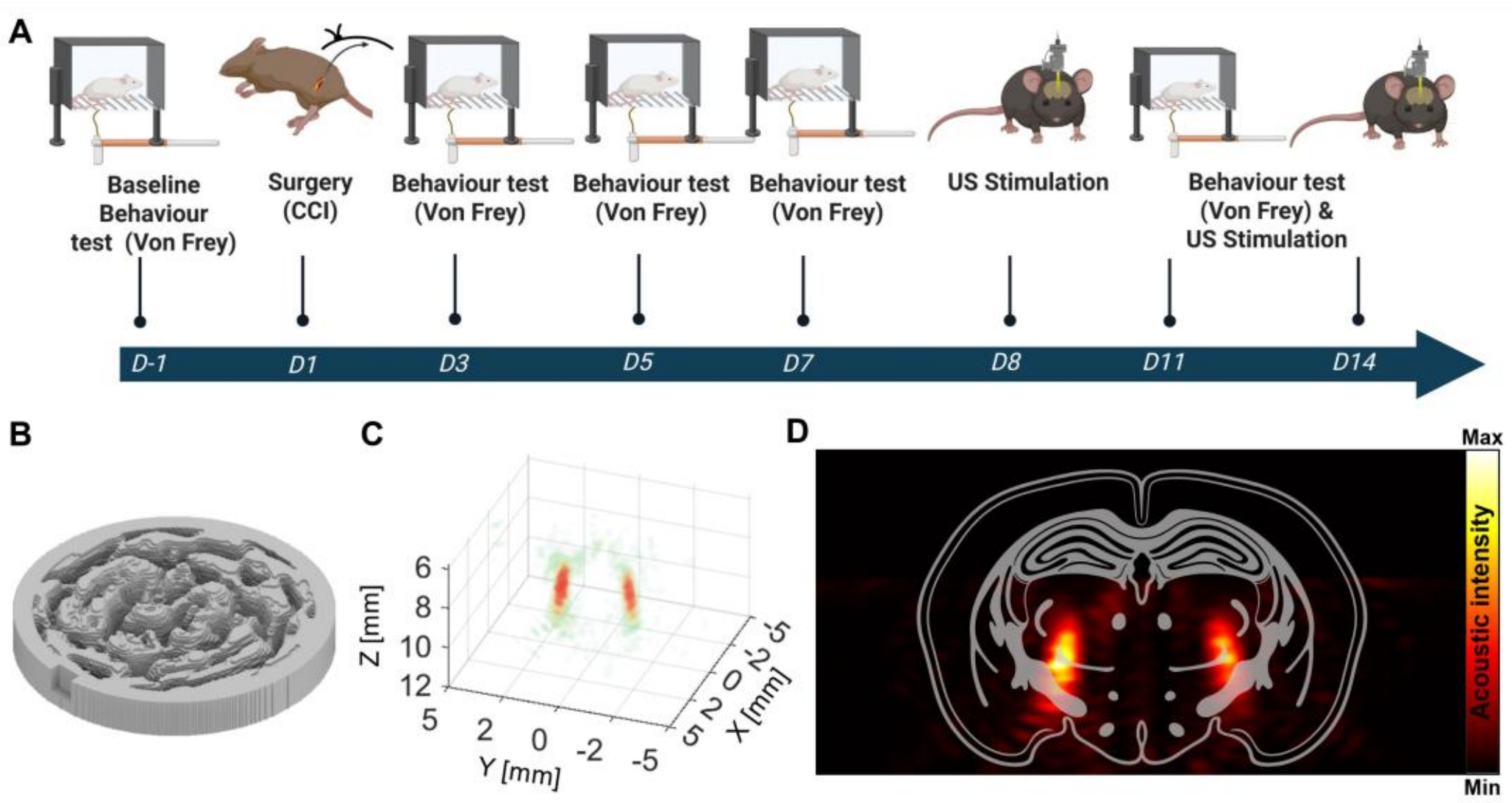

Figure 7: ***In vivo experimental design and multi-focal transcranial ultrasound targeting using TOAH.*** *(**A**) Experimental timeline of the in vivo study. (**B**) Rendering of the dual-foci TOAH lens designed for bilateral stimulation of the VP thalamic nuclei. (**C**) 3D reconstruction of the acoustic pressure field generated by the TOAH. (**D**) Overlay of the acoustic intensity distribution on a coronal mouse brain section referenced to the Allen Brain Atlas.*

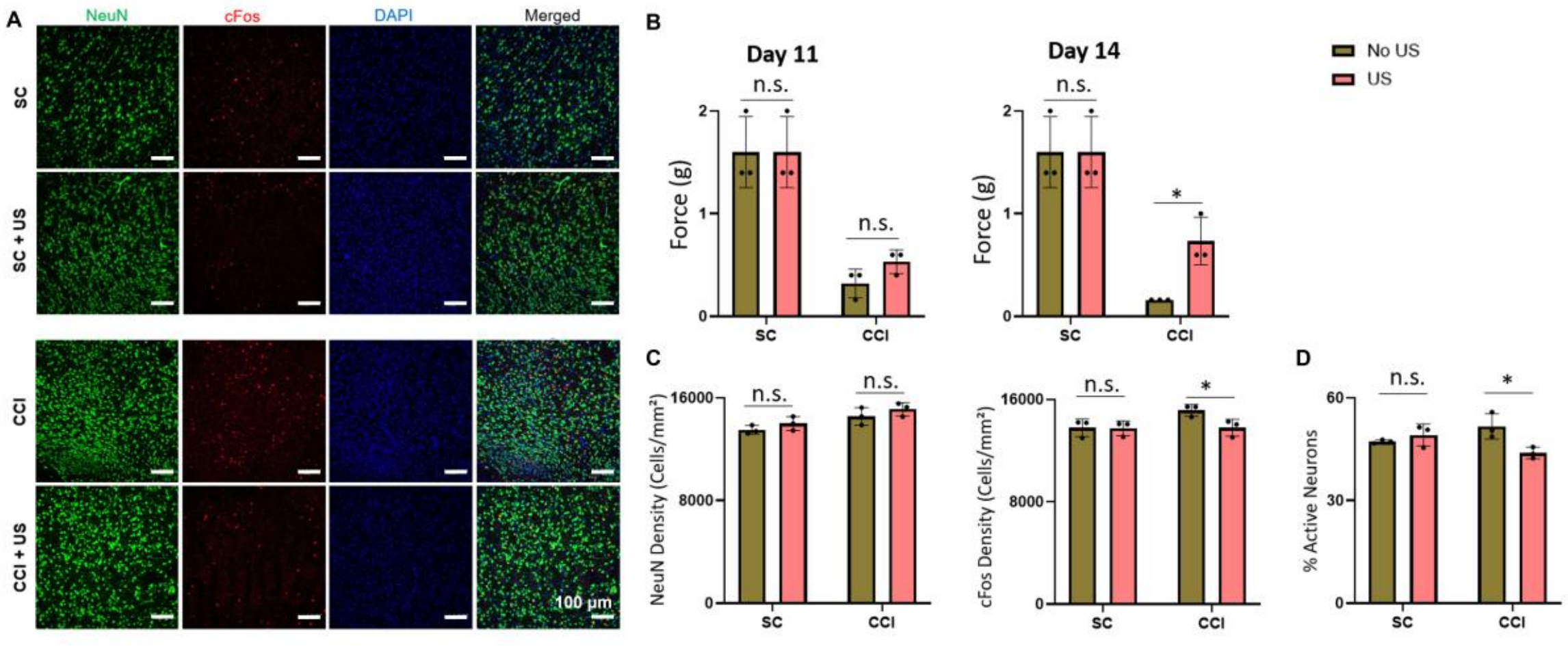

Figure 8: ***Behavioral and histological validation of TOAH-enabled multi-focal transcranial stimulation.*** *(**A**) Representative immunofluorescence images of VP nuclei showing NeuN (green), c-Fos (red), and DAPI (blue) staining for the SC, SC+US, CCI and CCI+US groups. (**B**) Mechanical withdrawal threshold measured using the von Frey test, demonstrating functional modulation following TOAH-enabled stimulation. (**C**) Quantification of NeuN$^+$ neuronal density (left) and c-Fos$^+$ cell density (right) in the VP nuclei. (**D**) Percentage of active neurons, defined as the ratio of c-Fos$^+$ to NeuN$^+$ cells. Data are presented as mean $\pm$ s.e.m. ($n = 3$ animals per group). $^*P < 0.05$, $^{**}P < 0.01$, and n.s., not significant; two-way ANOVA followed by Sidak's multiple-comparisons test. Dots represent individual animals.*

## In Silico Evaluation in Human Skull Models

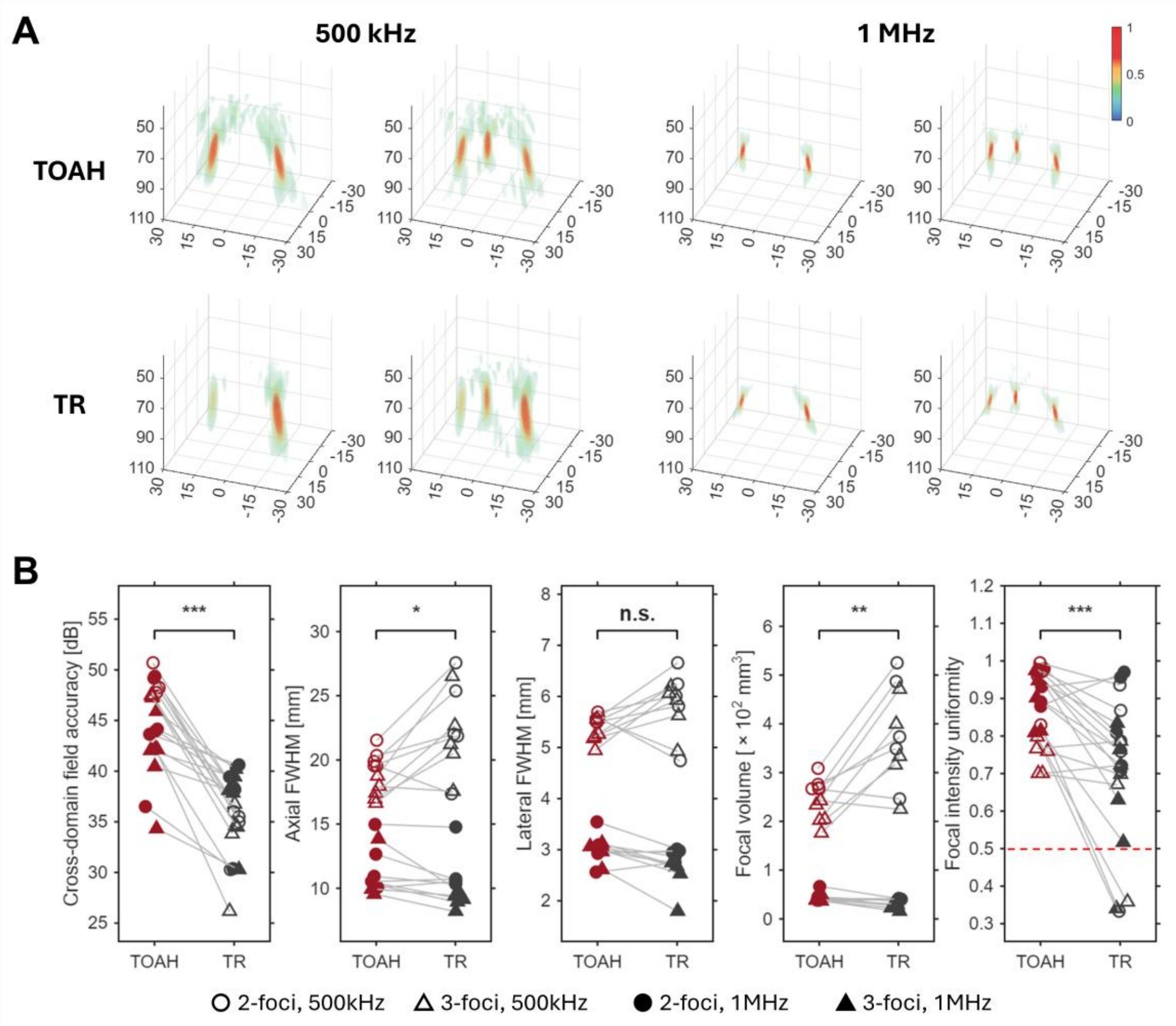

*Figure 9:* ***Human-skull simulation evaluation of TOAH across various skull models.*** *(**A**) Three-dimensional reconstructed acoustic pressure fields through one of the CT-derived human skull models (skull A) for 2-foci and 3-foci targeting at 500 kHz and 1 MHz. (**B**) Quantitative comparison of TOAH and TR across all matched skull–frequency–focal-configuration conditions (n=5). Metrics include cross-domain field accuracy, axial FWHM, lateral FWHM, focal volume, and focal intensity uniformity. Each connected pair represents the same skull, operating frequency, and focal configuration. Statistics reflect the overall TOAH-versus-TR effect from linear mixed-effects models. $^{*}P < 0.05$; $^{**}P < 0.01$; $^{***}P < 0.001$; n.s., not significant. The dashed red line denotes a focal intensity uniformity of 0.5.*

To assess the relative performance of TOAH under human-skull simulation conditions, additional in silico evaluations were conducted using CT-derived human skull models. The hologram was designed using TOAH to reconstruct dual-foci and tri-foci acoustic fields through the heterogeneous human skull. The performance was compared with a conventional TR-based hologram design approach, a representative baseline method for transcranial ultrasound hologram [12], [13], under identical acoustic and geometric conditions.

Figure 9A shows the simulated transcranial acoustic pressure fields reconstructed using TOAH and TR at $500 \text{ kHz}$ and $1 \text{ MHz}$. At $500 \text{ kHz}$, TR produced markedly imbalanced focal pressures, with one target receiving substantially higher pressure than the other. A similar imbalance was observed at $1 \text{ MHz}$, although the smaller focal regions made this difference less visually apparent. Quantitatively, Table 2 shows that, at $1 \text{ MHz}$, TOAH achieved a 25.7% higher mean focal intensity uniformity than TR. At the clinically relevant lower-frequency condition of $500 \text{ kHz}$, TOAH also produced more compact focal regions, with smaller axial and lateral FWHM values and a reduced focal volume relative to TR. At $1 \text{ MHz}$, TOAH showed comparable or modestly larger focal dimensions than TR, indicating a frequency-dependent trade-off between focal compactness and multifocal intensity balance.

To further assess robustness across anatomically variable human skull models ($n = 5$), TOAH and conventional TR were compared across matched skull, frequency, and focal-configuration conditions using linear mixed-effects models, with hologram design method, operating frequency, and number of foci specified as fixed effects and skull identity specified as a random effect. Hologram design method had significant effects on cross-domain field accuracy and focal intensity uniformity (Fig. 9B). Compared with TR, TOAH yielded an adjusted mean increase of 8.9 dB in cross-domain field accuracy and 0.170 in focal intensity uniformity (Table S4). Frequency-specific analyses further showed that TOAH improved cross-domain field accuracy and focal intensity uniformity by 38.2% and 22.3%, respectively, at $500 \text{ kHz}$ (Table 2). At $1 \text{ MHz}$, TOAH improved cross-domain field accuracy and focal intensity uniformity by 12.6% and 25.7%, respectively. Across matched conditions, TOAH maintained more consistent cross-domain field accuracy and focal intensity balance across skull models and focal configurations, supporting its robustness for multifocal acoustic field reconstruction under anatomically variable transcranial conditions.

**Table 2: Frequency-specific comparison of TOAH and TR across human skull models. Values are mean (standard deviation) across skulls and focal configurations.**

| Metric | 500 kHz | | 1 MHz | |
|---|---|---|---|---|
| | TOAH | TR | TOAH | TR |
| Cross-domain field-accuracy [dB] | 47.09 (2.78) | 34.08 (3.48) | 42.23 (4.34) | 37.49 (3.90) |
| Axial FWHM [mm] | 20.34 (4.13) | 24.26 (7.14) | 11.32 (1.85) | 10.25 (1.79) |
| Lateral FWHM [mm] | 5.40 (0.24) | 5.82 (0.59) | 2.99 (0.27) | 2.70 (0.36) |
| Focal volume [$\times 10^2 \text{ mm}^3$] | 2.45 (0.41) | 3.72 (1.00) | 0.43 (0.09) | 0.32 (0.09) |
| Focal intensity uniformity | 0.846 (0.119) | 0.692 (0.199) | 0.910 (0.062) | 0.724 (0.193) |

# Discussion

Acoustic holography has emerged as a promising approach for shaping acoustic fields and enabling flexible transcranial neuromodulation [13], [16], [19], [45], [46], [47]. Previous studies have demonstrated the feasibility of acoustic lenses and holographic approaches for transcranial phase aberration correction and shaping arbitrary acoustic fields. However, the performance of conventional hologram implementations remains limited by discrepancies between designed and physically delivered acoustic fields due to simplified phase-only approximations [20], [21], [48]. In these approaches, optimized phase distributions are typically converted into thickness-modulated physical lenses after optimization, which can introduce distortions in the realized acoustic field, particularly under complex multifocal transcranial conditions. To address this challenge, we introduce the physics-aware hologram technique, TOAH. It should be noted that the novelty of TOAH is not the use of an acoustic lens per se, but the direct optimization of the physically realizable three-dimensional thickness profile within the acoustic propagation mode. This design strategy reduces the mismatch between numerical design and physical implementation, leading to more accurate and reliable reconstruction of transcranial multifocal acoustic fields after fabrication. Furthermore, the physics-aware end-to-end optimization improves key focal characteristics, including focal volume and focal intensity uniformity. As a result, the generated acoustic fields can be delivered more consistently to the intended brain regions, enabling more reproducible neuromodulation outcomes.

This approach offers practical advantages in accessibility, implementation simplicity, and fabrication-consistent targeting over existing methods. While PhyAH [21] also optimizes the physical structure as TOAH, it requires specialized multi-material 3D printing and may introduce additional acoustic attenuation (Table S1). In contrast, the proposed TOAH maintains a thickness-varying plate architecture that can be readily implemented using standard single-material 3D printers, facilitating integration into existing experimental setups. In addition, TOAH provides a compact and cost-effective complementary alternative to high-element-count phased-array systems [19], [49] for multi-target transcranial neuromodulation. Although phased-array systems offer substantial flexibility through electronic beam steering and real-time focal control [50], they still require complex driving electronics, synchronization hardware, and multi-channel power amplification. By comparison, TOAH can generate complex transcranial multi-focal acoustic fields using a compact single-element transducer and a passive 3D-printed hologram lens. It is also noteworthy that the human-skull hologram consisted of 8,100 independently optimized elements, a scale that remains impractical for current phased-array ultrasound systems, which typically employ up to approximately 1,024 independently controllable elements. The simplified hardware configuration and fabrication accessibility of TOAH may facilitate broader adoption of transcranial multi-focal ultrasound technologies across diverse neuroscience and biomedical research settings.

We validated the TOAH through transcranial stimulation in rodents at 2-MHz, targeting deep brain structures with high spatial resolution (approximately 2 $mm$ axial and 700 $\mu m$ lateral FWHM). The strong agreement between ex vivo measurements and numerical simulations demonstrates that our approach improves both focal precision and the predictability of treatment planning, different from the previous studies [11], [12], [16]. This consistency is maintained even as target field complexity increases, indicating that the TOAH robustly supports simultaneous stimulation of distributed brain regions. Such capability is particularly relevant for neuroscience applications that require modulation of complex neural circuits or synergistic therapeutic effects through multi-site intervention [19], [51], suggesting TOAH as a promising platform for precision non-invasive brain stimulation.

Based on the in silico and ex vivo results, we extended the evaluation to in vivo experiments. The in vivo brain stimulation platform builds upon previously developed stereotaxic-guided ultrasound systems designed for precise targeting [37], [38]. Building on this foundation, our approach enables precise multi-focal stimulation using a single-element ultrasound transducer, extending the capabilities of existing platforms. Furthermore, leveraging the same underlying principles, the proposed technique can be extended to wearable implementations, as demonstrated in recent studies for freely moving behavioral experiments [4], [52], [53].

The in vivo study was not designed to establish therapeutic efficacy but rather to demonstrate reliable and spatially localized in vivo delivery enabled by TOAH. Nevertheless, we observed outcomes consistent with potential therapeutic effects. Specifically, we applied TOAH to a chronic neuropathic pain model by targeting the bilateral VP nuclei, a clinically established target for deep brain stimulation [43], [44]. Using continuous theta-burst ultrasound stimulation (cTBUS) parameters adapted from prior work [39], the treated group exhibited reduced c-Fos expression within the targeted nuclei, accompanied by a significant increase in mechanical withdrawal thresholds. These findings support that TOAH enables functionally relevant and multi-focal transcranial ultrasound neuromodulation in vivo. While the observed behavioral and cellular responses are consistent with prior reports [39], the present study differs in both anatomical target and stimulation configuration, and therefore does not directly establish equivalent mechanisms. The reduction in c-Fos expression suggests modulation of neuronal activity within the targeted thalamic regions; however, the precise neurobiological mechanisms underlying this effect remain to be determined. In addition, the in vivo experiments were performed with a limited sample size ($n = 3$ per group), and the behavioral findings should therefore be interpreted as preliminary proof-of-concept results. Future studies with larger cohorts will be necessary to further evaluate the reproducibility, effect size, and therapeutic significance of TOAH-enabled neuromodulation. Overall, these results highlight the capability of TOAH to deliver controlled and spatially precise multi-focal stimulation through the intact skull, while also motivating future investigations into its potential therapeutic applications.

Although the proposed TOAH framework demonstrated robust transcranial multi-focal reconstruction through the rodent skull, the acoustic properties and geometric complexity of the human skull differ substantially from those of rodents. In particular, its greater thickness, heterogeneity, and attenuation can introduce more severe phase aberration and energy loss, potentially affecting focusing performance in practical human applications. To partially address this limitation, we additionally performed in silico evaluations using human skull models and compared TOAH with the conventional TR-based approach. Although CT-derived acoustic-property assignments involve inherent uncertainty, previous experimental–simulation comparisons have shown that such simulations can reasonably predict focal location, focal size, and overall spatial field distribution despite quantitative uncertainty in absolute pressure estimates [40]. Accordingly, the present human-skull simulations were used to assess relative spatial field reconstruction and focal precision, rather than absolute pressure-dose accuracy, consistent with their use in transcranial ultrasound treatment planning and focal-targeting evaluation [30], [54].

Under identical human-skull simulation conditions, TOAH achieved higher cross-domain field fidelity and substantially improved focal intensity uniformity compared with TR. Improved focal intensity uniformity is particularly important for multi-focal stimulation because pressure imbalance across target regions can result in nonuniform stimulation effects. Such imbalance has been recognized as a major limitation of conventional TR-based transcranial multi-focal reconstruction [12] and was also observed in the present study for TR-based hologram design (Fig. 9A). When comparing the overall focal dimensions across operating frequencies, 1 MHz provided improved spatial confinement relative to 500 kHz, consistent with its shorter acoustic wavelength and reduced diffraction-limited spreading. However, 1 MHz remains less commonly used in transcranial ultrasound applications, and its translational feasibility requires further investigation. In particular, the increased skull attenuation at higher frequencies may reduce transmitted acoustic energy while increasing the potential for skull heating and associated safety concerns.

Regarding focal confinement, which is also important for spatially precise stimulation, frequency-specific differences were observed. At $500\ \mathrm{kHz}$, TOAH produced a smaller focal volume than TR. At $1\ \mathrm{MHz}$, however, TOAH showed comparable or modestly larger focal dimensions than TR. This pattern may reflect the stronger skull-induced phase aberration at higher frequencies, which makes simultaneous balancing of multiple focal pressures more challenging. Under these conditions, TOAH may prioritize balanced energy delivery across the intended targets, resulting in a trade-off between focal intensity uniformity and focal compactness. A similar trade-off was observed in the rodent-skull simulations performed at $2\ \mathrm{MHz}$. These findings suggest that, at clinically relevant lower frequencies such as $500\ \mathrm{kHz}$, TOAH can support spatially confined and well-balanced multi-focal stimulation in the human brain. In contrast, for preclinical rodent studies, where higher frequencies are often required because of the smaller brain volume, TOAH may provide a useful strategy for maintaining balanced multi-focal energy delivery while preserving an adequate focal volume. It should note that different weight of loss terms also can guide the hologram optimization objective for more balanced energy or higher precision. Importantly, this balance–compactness trade-off can be adjusted during hologram optimization by varying the relative weights of the field reconstruction accuracy, focal-energy balance, and energy efficiency terms in the loss function, depending on whether a given application prioritizes uniform multi-target energy delivery, higher focal pressure with a limited aperture, or maximal spatial confinement.

Collectively, these CT-derived simulations suggest that TOAH may improve multifocal spatial field reconstruction relative to conventional TR-based hologram design under human transcranial conditions. However, these findings should be interpreted as preliminary simulation-based evidence rather than definitive validation of human application. Experimental measurements using human skull specimens will be important to further verify the reconstructed pressure fields and to calibrate in situ pressure delivery and absolute acoustic dose before translational application.

Another limitation of the current TOAH framework is that it is formulated for planar aperture geometries, whereas previous human transcranial ultrasound studies have commonly employed spherically curved apertures to achieve lower f-numbers and improved axial focusing performance. Incorporating curved-aperture configurations into the TOAH framework may therefore further improve spatial confinement. Extending the present acoustic wave propagation model to support curved-aperture propagation will require additional methodological development and will be investigated in future studies.

In conclusion, we present a physics-aware hologram technique, thickness-only acoustic hologram, that reduces the numerical–physical gap inherent in conventional approaches, enabling more reliable and accurate transcranial ultrasound neuromodulation. By directly optimizing the three-dimensional physical design, the proposed method achieves consistent and balanced multi-region ultrasonic focusing under complex transcranial conditions. Through comprehensive validation—including in silico and ex vivo evaluations and in vivo studies in a rodent model—we demonstrate improved focal precision and pressure-balance and simulation-to-physical fidelity, along with reproducible neuromodulatory effects. Although the additional human-skull evaluation was limited to in silico analysis, the results provide preliminary simulation-based evidence that the relative reconstruction advantages of TOAH may extend to challenging human transcranial conditions. These results establish a practical and robust approach for high-precision, non-invasive brain stimulation and highlight its potential for future neuroscience and neuromodulation research.

# Acknowledgments

*Funding:*

This work was partly by the National Research Foundation of Korea(NRF) grant funded by the Korea government (Ministry of Science and ICT, MSIT) under Grant RS-2024-00354020 and partly by the National Research Foundation of Korea (NRF) grant funded by Korean Government (Ministry of Science and ICT, MSIT) under Grant RS-2023-00302281.

*Author contributions:*

Conceptualization: M.H.L., and J.Y.H. Methodology: M.H.L., and M.A.K. Investigation: M.H.L., M.A.K., A.A., E.L., J.L., E.C., H.-S.K., and J.Y.H Visualization: M.H.L., and M.A.K. Resources: E.C., H.-S.K., and J.Y.H. Supervision: E.C., H.-S.K., and J.Y.H. Writing—original draft: M.H.L., M.A.K. and J.Y.H. Writing—review and editing: M.H.L., E.C., H.-S.K., and J.Y.H.

*Competing interests:*

There are no competing interests to declare.

*Data and materials availability:*

All data needed to evaluate the conclusions in the paper are present in the paper and/or the Supplementary Materials.

# Supplementary materials

Supplementary Sections S1 to S5
Figs. S1 to S11
Tables S1 to S4

# Supplementary Materials for High-Fidelity Transcranial Ultrasound Multi-focal Stimulation via Physics-Aware Hologram Technique

Moon Hwan Lee, Mohd. Afzal Khan, Akm Ashiquzzaman, Eunbin Lee, Jonghun Lee, Euiheon Chung*, Hyuk-Sang Kwon*, Jae Youn Hwang*
*Corresponding author. Email: ogong50@gist.ac.kr, hyuksang@gist.ac.kr, jyhwang@dgist.ac.kr

## This PDF file includes:

Supplementary Sections S1 to S5
Figures S1 to S11
Tables S1 to S4

## Supplementary Text

### Section S1. Overview of Phase-Only Acoustic Hologram (POAH)

Currently, almost all the application studies of acoustic hologram are using the POAH. The POAH design process proceeds as follows: First, the uniform acoustic pressure field emitted by an ultrasound transducer is modulated by a phase delay map, denoted by $\phi(i, j)$, where $i$ and $j$ index the spatial coordinates. Mathematically, the source pressure is expressed as

$$P_{\text{source}} = A_{\text{source}} e^{j\phi},$$

where $P_{\text{source}}$ and $A_{\text{source}}$ represent the complex pressure field and amplitude distribution at the source plane, respectively. The propagation of acoustic waves is modeled by an operator $f$, such that the simulated acoustic pressure field is given by the following equation:

$$P_{\text{sim}} = f(P_{\text{source}}).$$

Accordingly, the design objective is to find an optimal phase map $\phi^*$ that minimizes the mismatch between the simulated acoustic field $P_{\text{sim}}$ and the desired target field $P_{\text{target}}$:

$$\phi^* = \underset{\phi}{\operatorname{argmin}};\mathcal{L}\left(P_{\text{sim}}, P_{\text{target}}\right),$$

where $\mathcal{L}$ measures the discrepancy between the simulated and target fields.

Typically, this inverse problem has been solved using the iterative angular spectrum approach (IASA)[10], while time reversal (TR) techniques are employed for heterogeneous media setups. More recently, gradient-based optimization has been introduced [55]. For the gradient-based optimization, applying the chain rule, the gradient of the loss with respect to the phase is expressed as

$$\frac{\partial \mathcal{L}}{\partial \phi} = \frac{\partial \mathcal{L}}{\partial f} \frac{\partial f}{\partial \phi},$$

where both $\frac{\partial \mathcal{L}}{\partial f}$ and $\frac{\partial f}{\partial \phi}$ are available as long as the loss function $\mathcal{L}$ and propagation operator $f$ are differentiable.

For fabrication, the optimized phase map must be converted into a thickness profile $T(i, j)$:

$$T(i, j) = \frac{\theta(i, j)}{2\pi f \left(\frac{1}{c_0} - \frac{1}{c_L}\right)},$$

where $f$ is the operating frequency of the ultrasound transducer, $c_0$ is the sound speed of the surrounding medium (e.g. water or tissue), and $c_L$ is the sound speed of the 3D-printing material. The resulting thickness profile is converted to 3D geometry annd then fabricated via 3D printing to form the hologram lens.

## Section S2. Supplementary notes on DHLA

**Algorithm S1** Differentiable Hologram Lens Approximation (DHLA)

**function** $\Psi(\theta, V_{\max}, V_{\min}, \alpha, \beta)$

**Require:** $\theta \in \mathbb{R}^{N_x \times N_y}$

**Parameters:**

$V_{\max}$: Maximum lens thickness (in voxels)

$V_{\min}$: Minimum lens thickness (in voxels)

$\alpha$: Sigmoid steepness for thickness mapping

$\beta$: Sigmoid steepness for voxel transition

**Ensure:** $V' \in [0,1]^{N_x \times N_y \times V_{\max}}$ ▹ Quasi-binary 3D volume

$t(i,j) \leftarrow \sigma(\alpha \cdot \theta(i,j)) \cdot (V_{\max} - V_{\min}) + V_{\min}$

▹ Map each 2D point to continuous thickness

$t_{\text{smooth}}(i,j) \leftarrow (G * t)(i,j)$ where $G$ is a $9 \times 9$ Gaussian kernel

▹ Apply spatial smoothing filter via convolution

**for** $k = 1$ to $V_{\max}$ **do**

$V'(i,j,k) \leftarrow \sigma(\beta \cdot (k - t_{\text{smooth}}(i,j)))$ for all $i \in [1, N_x], j \in [1, N_y]$

▹ Generate soft voxel occupancy along the thickness dimension

**end for**

**return** $V'$

**end function**

In this study, $\alpha$ was set to 0.1 and the value of $\beta$ is initialized at 1 and gradually annealed to 0.1 during optimization, steepening the 0–1 transition and progressively pushing the voxel representation toward hard binarization.

## Section S3. Supplementary notes on the TOAH frameowork

**Media Adjustment and Lens Embedding:** The heterogeneous medium $\boldsymbol{M}$ is represented over a 3D grid, where each voxel $M_{ijk}$ stores its acoustic properties:

$$\boldsymbol{M} = M_{ijk} \mid i \in [1, N_x], j \in [1, N_y], k \in [1, N_z]$$

$$M_{ijk} = (c_{ijk}, \rho_{ijk}, \alpha_{ijk})$$

with $c_{ijk}$ (m/s), $\rho_{ijk}$ (kg/m$^3$), and $\alpha_{ijk}$ $(\mathrm{dB}/(\mathrm{MHz}^y \cdot \mathrm{cm}))$ representing the speed of sound, density, and attenuation coefficient at voxel $(i,j,k)$, respectively. The grid dimensions $N_x$, $N_y$, $N_z$ and spacings $d_x$, $d_y$, $d_z$ define the simulation domain size and resolution, respectively.

The lens geometry is embedded into the wave-propagation model by modifying $\boldsymbol{M}$ as follows:

1. Represent the lens as a quasi-binary volume

$$V' \in [0,1]^{N_x \times N_y \times N_v},$$

   where $N_v$ denotes the maximum lens thickness in grid units. Voxels near 1 correspond to lens material; those near 0 correspond to the background medium.

2. Construct three property volumes $V_c', V_\rho', V_\alpha'$ by element-wise multiplying $V'$ with the speed-of-sound, density, and attenuation coefficient of the lens material, respectively.

3. For each voxel where $V'_{ijk} > 0.9$, replace

$$(c_{ijk}, \rho_{ijk}, \alpha_{ijk}) \quad \text{with} \quad (V'_{c,ijk}, V'_{\rho,ijk}, V'_{\alpha,ijk}).$$

**Optimization Objective:** We construct the total loss as

$$\mathcal{L} = \mathcal{L}_{\text{acc}} + \lambda_{\text{energy}} \cdot \mathcal{L}_{\text{energy}} + \lambda_{\text{balance}} \cdot \mathcal{L}_{\text{balance}}.$$

We optimize the acoustic pressure field with respect to amplitude; thus, the target acoustic field is represented by $A_{target}$. The loss functions are defined as follows. The weighting parameters $\lambda_{\text{energy}}$ and $\lambda_{\text{balance}}$ are experimentally tuned to balance performance.

1. **Field reconstruction accuracy ($\mathcal{L}_{\text{acc}}$).** The accuracy term is computed on intensity rather than pressure amplitude, which amplifies discrepancies in high-pressure regions:

$$\mathcal{L}_{\text{acc}} = 1 - \frac{\sum_{x=1}^{N_x}\sum_{y=1}^{N_y}\sum_{z=1}^{N_z} A_{\text{target}}(x,y,z)^2 \, |P_{\text{sim}}(x,y,z)|^2}{\sqrt{\left(\sum_{x=1}^{N_x}\sum_{y=1}^{N_y}\sum_{z=1}^{N_z} A_{\text{target}}(x,y,z)^4\right)\left(\sum_{x=1}^{N_x}\sum_{y=1}^{N_y}\sum_{z=1}^{N_z} |P_{\text{sim}}(x,y,z)|^4\right)}}.$$

2. **Energy efficiency ($\mathcal{L}_{\text{energy}}$).** Because of the limited aperture size, the total available energy is constrained; therefore, it is desirable to allocate more energy to the target regions. The energy-efficiency term maximizes the average intensity within the desired focal region defined by $A_{\text{target}}$:

$$\mathcal{L}_{\text{energy}} = -\frac{\sum_{x=1}^{N_x}\sum_{y=1}^{N_y}\sum_{z=1}^{N_z} A_{\text{target}}(x,y,z) \cdot |P_{\text{sim}}(x,y,z)|}{\sum_{x,y,z} A_{\text{target}}(x,y,z)},$$

3. **Focal-energy balance ($\mathcal{L}_{\text{balance}}$).** We aim to achieve a uniform acoustic pressure across the target regions. To ensure uniform acoustic energy across multiple focal points, the balance term minimizes the intensity variation inside the active target region $\Omega$:

$$\mathcal{L}_{\text{balance}} = \text{Std}\left(\left\{A_{\text{target}}(x,y,z)\,|P_{\text{sim}}(x,y,z)|^2\right\}_{(x,y,z)\in\Omega}\right),$$

$$\Omega = \left\{(x,y,z)\;\middle|\; A_{\text{target}}(x,y,z) = 1\right\}.$$

**Algorithm implementation:** The hologram optimization pipeline (Algorithm S2), including the DHLA ($\Psi$) and the mixed-domain wave solver, was implemented in Python using TensorFlow to fully exploit automatic differentiation and GPU acceleration. The reflection order of the wave solver was set to 4, allowing up to fourth-order reflections, which have shown to be sufficient for stabilizing simulation errors [23].

The hologram geometry was optimized for 200 iterations using the Adam optimizer [56] with a learning rate of 1, selected empirically. Regarding the loss, we set $\lambda_{\text{energy}} = 2\times10^{-1}$, $\lambda_{\text{balance}} = 5\times10^{-1}$. All computations were performed on a workstation equipped with an Intel i7-9800X CPU, 64 GB RAM, and an NVIDIA RTX 3090 Ti GPU. The implementation required approximately 9.7 GB of GPU memory, and each optimization iteration took approximately 3 s, corresponding to a total runtime of about 10 min for 200 iterations.

**Algorithm S2** Gradient Descent Optimization of V'

**Require:** Transducer parameters $\mathbf{U}$, target heterogeneous medium $\mathbf{M}$, initial 2D parameter map $\theta_0$, learning rate $\eta$, number of iterations $N$, target amplitude $A_{target}$

**Ensure:** Optimized 3D quasi-binary lens matrix $V'_{opt}$

Initialize $\theta \leftarrow \theta_0$

**for** $i \leftarrow 1$ to $N$ **do**

  $V' \leftarrow \Psi(\theta)$ ▷ Apply DHLA (Algorithm S1)

  $\mathbf{M} \leftarrow \text{UpdateMediaProperties}(V')$ ▷ Embed lens into heterogeneous media

  $P_{sim} \leftarrow f(\mathbf{U},\mathbf{M})$ ▷ Simulate acoustic field using mSOUND

  $\mathcal{L} \leftarrow \text{ComputeLoss}(|P_{sim}|, A_{target})$ ▷ Compute loss

  $\nabla_\theta \mathcal{L} \leftarrow \frac{\partial \mathcal{L}}{\partial V'} \cdot \frac{\partial V'}{\partial \theta}$ ▷ Compute gradient

  $\theta \leftarrow \theta - \eta \nabla_\theta \mathcal{L}$ ▷ Gradient descent

**end for**

$V'_{opt} \leftarrow \Psi(\theta)$ ▷ Obtain final optimized V'

**return** $V'_{opt}$

## Section S4. Supplementary notes on the ex-vivo acoustic pressure measurement

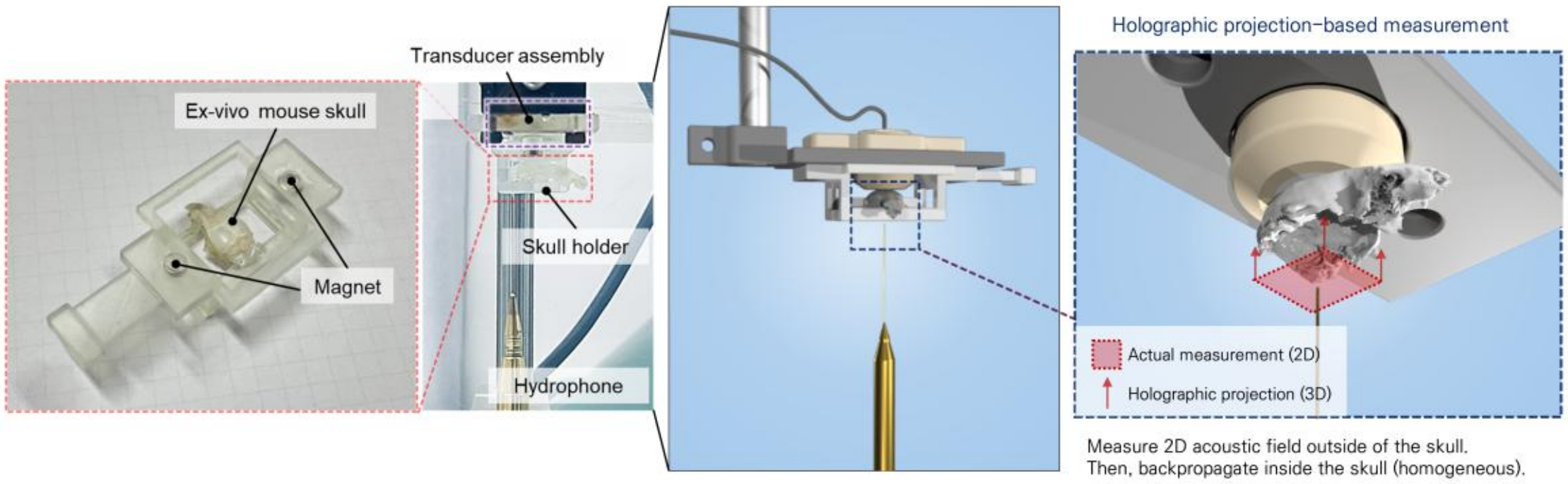

*Figure S1: **Ex-vivo measurement setup**. Skull holder is loaded with a harvested mouse skull. The skull holder is fixed beneath the transducer mount, while the transducer assembly is secured above; the hydrophone is translated by a 3-axis stage to map the acoustic field.*

Fig. S1 shows the experimental setup for the ex-vivo acoustic pressure measurement. A mouse skull was mounted on a custom-designed fixture considering its micro-CT scan to replicate the transducer–skull geometry used in simulation.

**Calibration procedure:** For controlled experiments, we designed and fabricated a calibration hologram lens that generates a single on-axis focus at a depth of 8 mm in water (Fig. S2). Using this lens, we experimentally identified the location of maximum pressure, 8 mm from the transducer aperture, and defined it as the origin for all subsequent measurements. Since the relative positions of the transducer, hologram, and skull are known in the simulation, this origin was used as a reference to align the simulation with the experimental measurements.

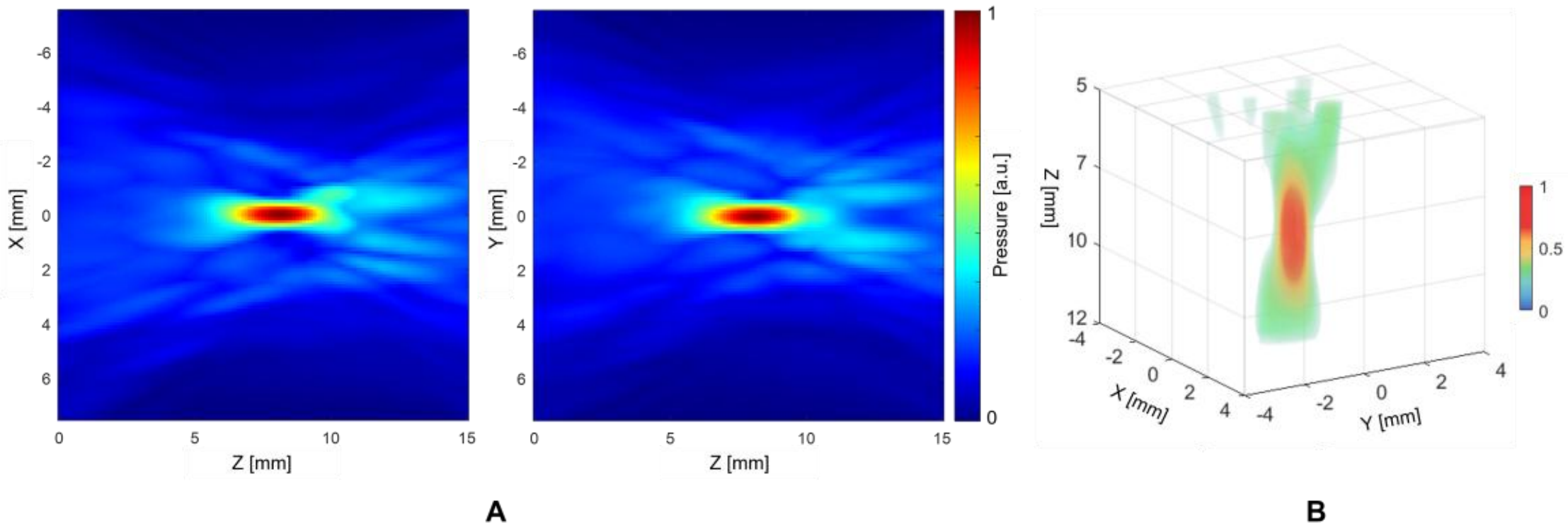

*Figure S2: **Acoustic field measurement of the calibration hologram lens**. (**A**) XZ and YZ cross-sections of the acoustic field and (**B**) the 3D reconstructed acoustic field.*

**Hydrophone scanning parameters and data acquisition protocol:** Because the ex-vivo skull lacks the inferior boundary for hydrophone measurement, it is not possible to account for multiple internal reflections within the cranial cavity as assumed in the simulation design. Nevertheless, as shown in Fig. S5, a pulse length of 5 to 10 cycles, which is shorter than the time required for even the first reflection to arrive, is sufficient to confirm the acoustic field reconstruction performance. Therefore, a 10-cycle pulse was used for all ex-vivo measurements.

For each measurement:

- The hydrophone was positioned 3 mm below the reference origin plane.
- A raster scan was performed over a 13.65 mm × 13.65 mm region.
- Spatial sampling step: 150 $\mu\text{m}$.
- Four acquisitions were averaged per spatial point for consistent measurement.
- Driving waveform: 2 MHz sinusoid, 5 kHz PRF, 10 cycles per pulse.

These settings ensured sufficient spatial resolution for reconstructing the target field.

**Reconstruction of 3D fields via angular spectrum method:** Acoustic signals recorded at each scan point were processed using the `extractAmpPhase` function in the k-Wave toolbox [29], yielding the complex pressure at the driving frequency. The full 3D field was reconstructed by backpropagating this complex field using the angular spectrum method. This reconstruction relies on the assumption of a homogeneous medium beneath the skull—a condition satisfied in the experimental setup.

## Section S5. Thermal simulations

Thermal simulations were done by solving the Pennes bioheat transfer equation using `kWaveDiffusion` function of the k-Wave toolbox [29]. We performed the thermal simulations on the same voxelized domain used for acoustic modeling. Voxel-wise thermal conductivity and specific heat capacity were assigned based on tissue type and density: for bone, k=0.32 $\text{W m}^{-1}\,^{\circ}\text{C}^{-1}$ and c=1313 $\text{J kg}^{-1}\,^{\circ}\text{C}^{-1}$; for soft tissue, k=0.51 $\text{W m}^{-1}\,^{\circ}\text{C}^{-1}$ and c=3630 $\text{J kg}^{-1}\,^{\circ}\text{C}^{-1}$ [50]. We simulated five cycles of heating (10 ms at peak pressure) and cooling (190 ms), corresponding to 1 sec at which already showed temperature rise nearly 3°C. The sonication protocol was selected to match representative in vivo blood–brain barrier opening experiments (5 Hz pulse repetition frequency, 10 ms pulse duration), which typically require longer exposure durations than those used in neuromodulation studies. To isolate the effect of design-dependent pressure variations, we normalized each hologram's acoustic field such that the negative peak pressure within the target region was 1 MPa prior to thermal simulation. It corresponds to 1.7 $W/cm^2$ of $I_{spta}$, 0.71 of mechanical index.

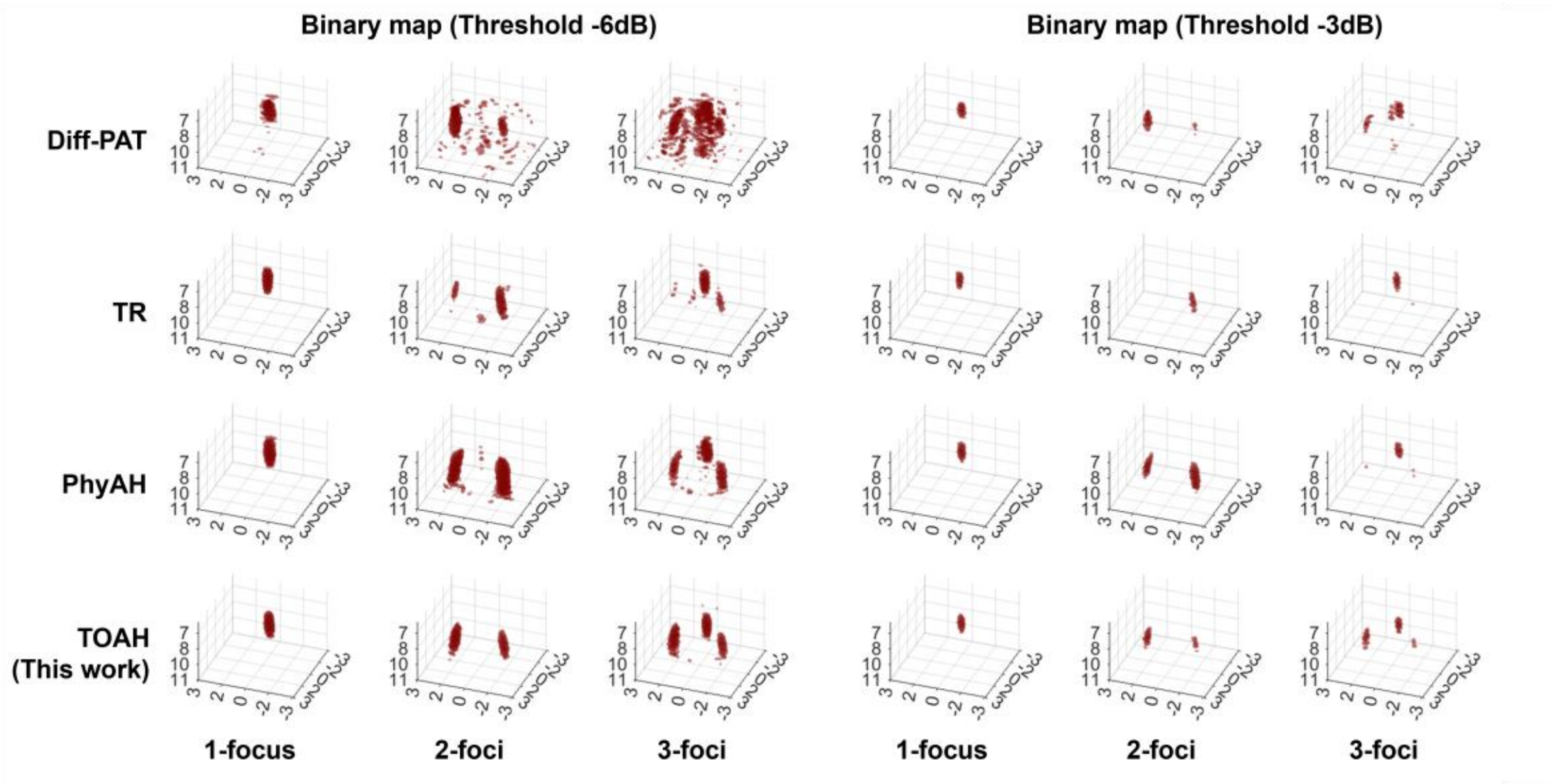

*Figure S3:* ***Three-dimensional binarized pressure fields in the fabrication domain for single-, dual-, and tri-focal configurations using -3 dB and -6 dB thresholds.***

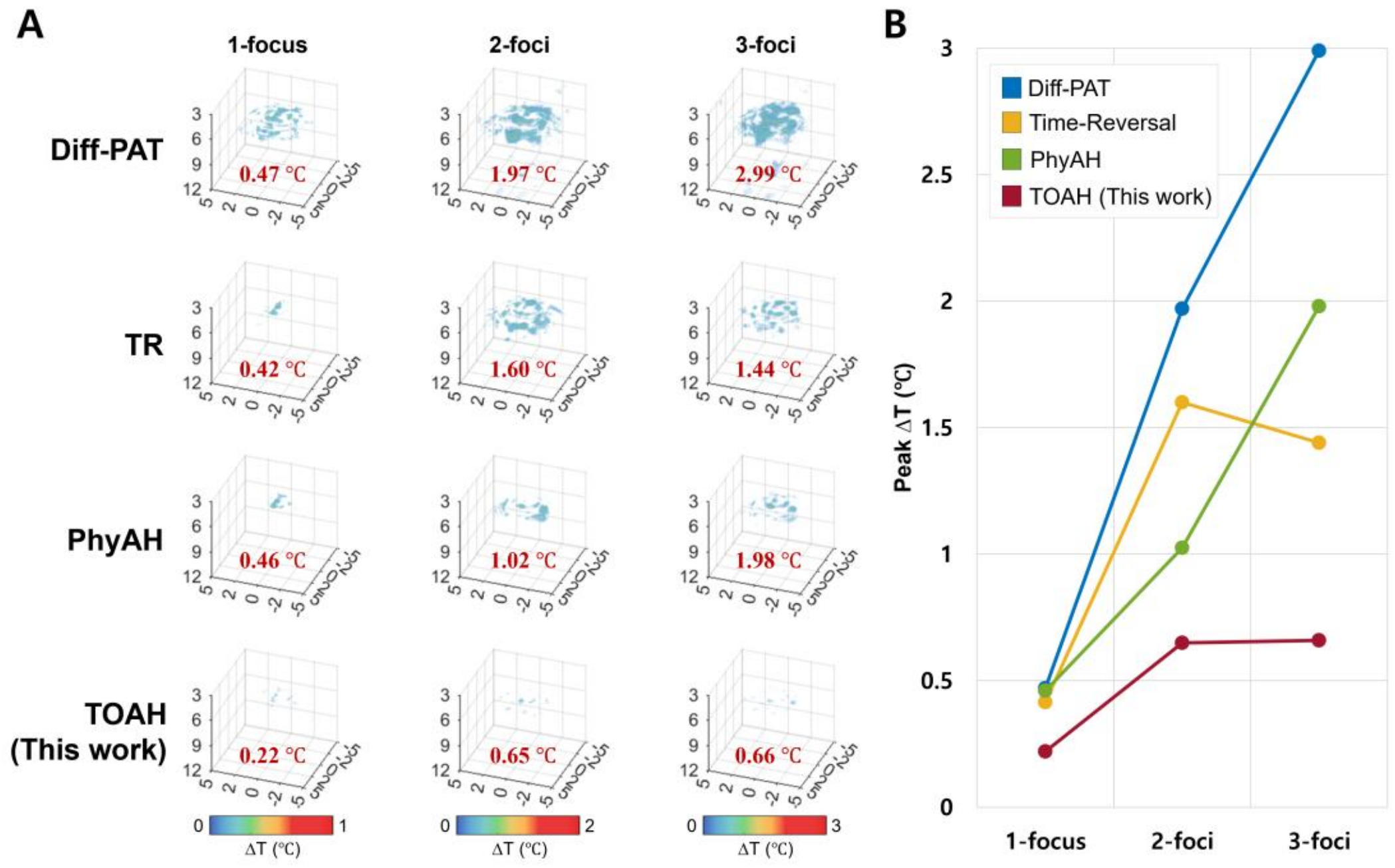

*Figure S4:* ***Thermally induced temperature rise (*** $\Delta T$ ***) in the skull for different acoustic hologram design methods.*** *(**A**) Three-dimensional temperature distributions obtained from thermal simulations for single-, dual-, and tri-focal excitation. Phase-based methods exhibit increased off-target heating as the number of foci increases, whereas TOAH maintains localized and reduced temperature elevation. (**B**) Peak temperature rise within the skull for each focusing configuration. The TOAH maintains a temperature increase below* $1^{\circ}C$ *even as the number of foci increases.*

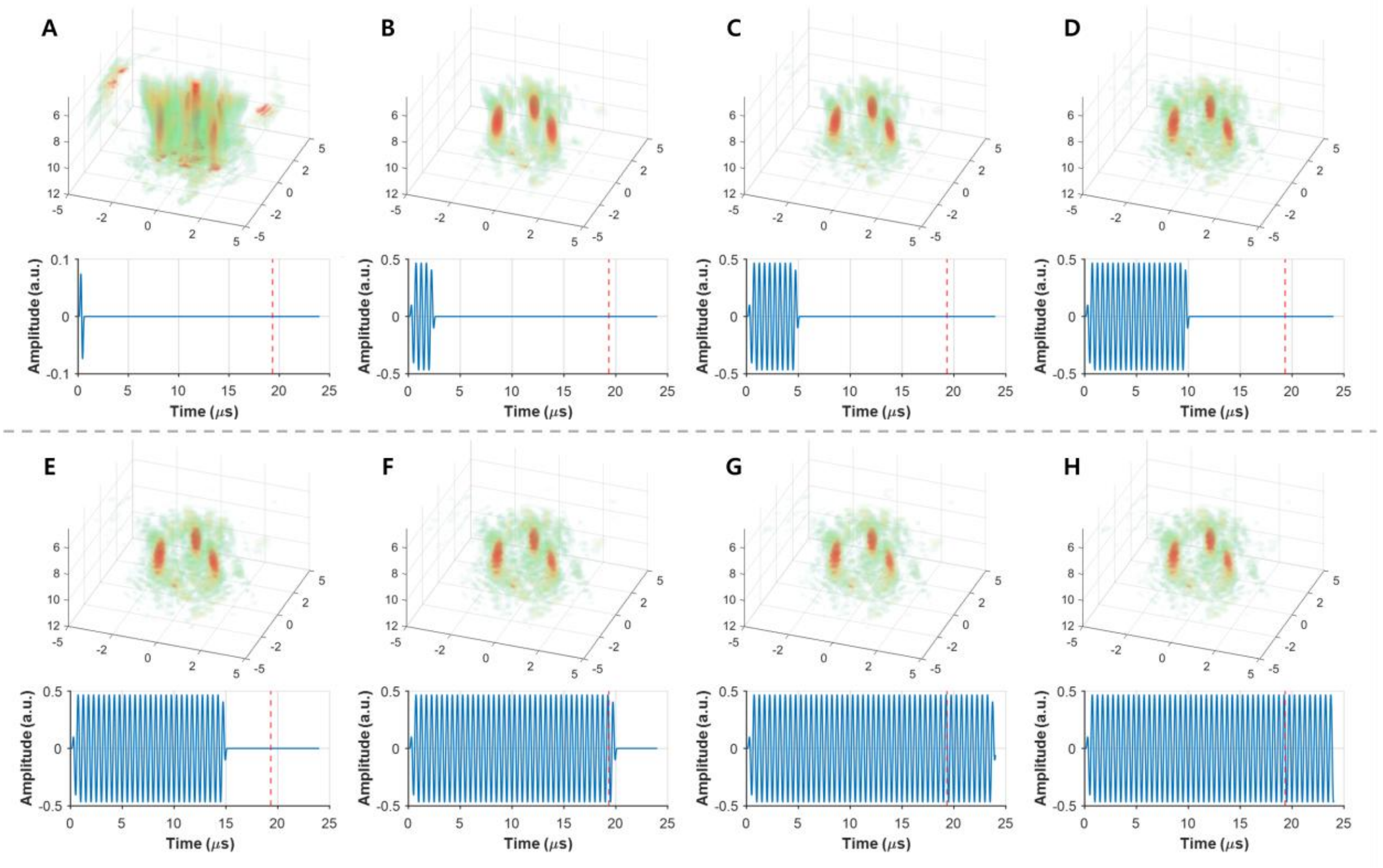

*Figure S5:* ***Simulated three-dimensional pressure fields (top row) and corresponding driving waveforms (bottom row) for increasing pulse lengths from 1 cycle to continuous wave (CW).*** *The red dashed line indicates the arrival time of the fourth-order reflection in the time-domain simulation. After 5 cycles, the reconstructed fields begin to resemble the target pattern, although standing-wave effects gradually appear as the pulse length increases, producing spatially oscillatory artifacts across the focal region.*

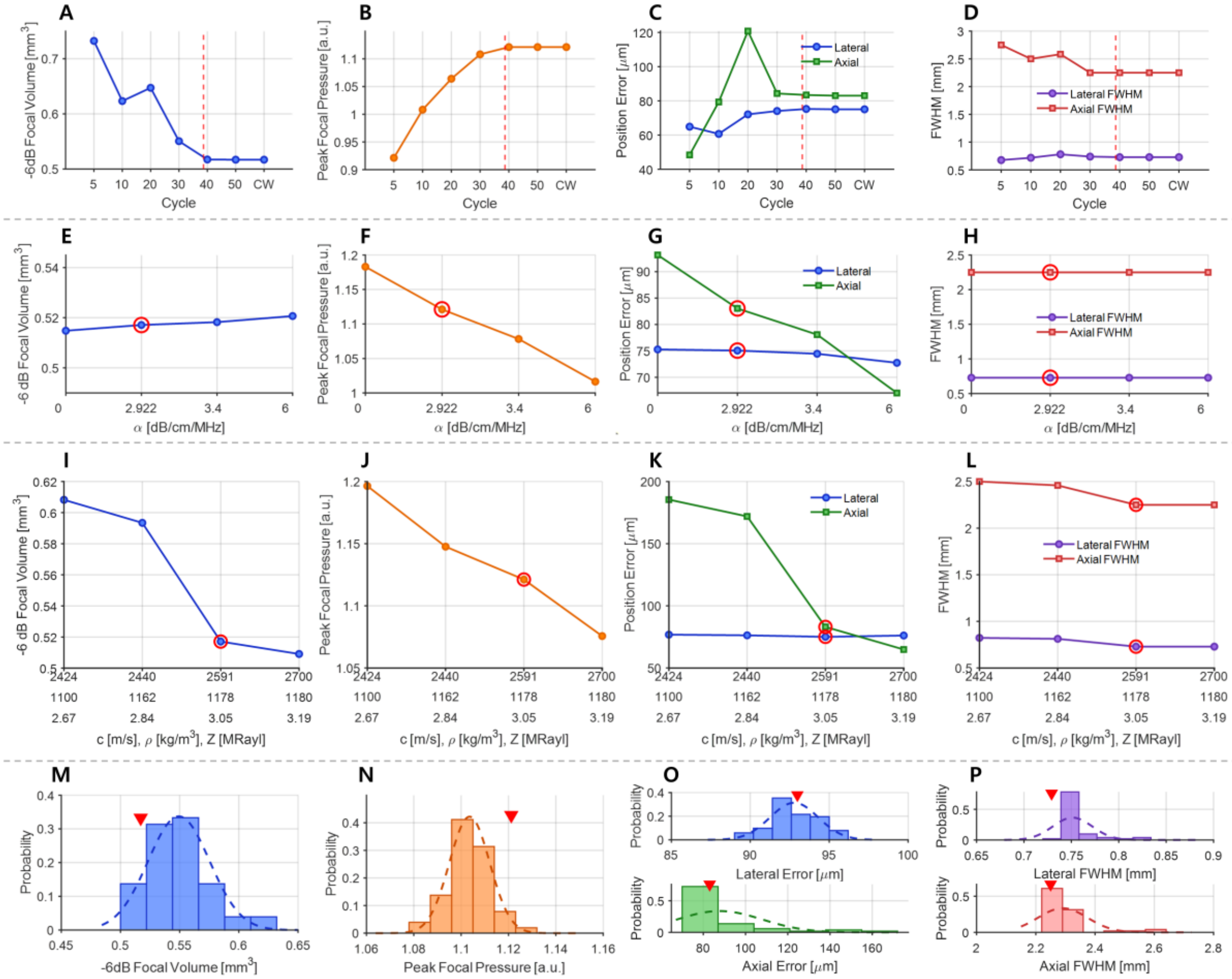

Figure S6: **Sensitivity analysis of acoustic field reconstruction performance with respect to pulse length, lens material properties, and fabrication-induced surface deviations.** (**A**–**D**) Dependence of key quantitative metrics on pulse length, including –6 dB focal volume, peak focal pressure, lateral and axial target position errors, and focal spot size quantified by the lateral and axial FWHM. Red dashed lines indicate the arrival of the fourth-order reflection in the time-domain simulation. (**E**–**L**) Effect of acoustic attenuation, sound speed and density of the lens material on field reconstruction performance. Red circles indicate the acoustic property values used in the proposed hologram design. (**M**–**P**) Statistical distributions of reconstruction metrics obtained from 50 independent realizations with random thickness perturbations applied to the lens. Red triangles denote the performance of the unperturbed design.

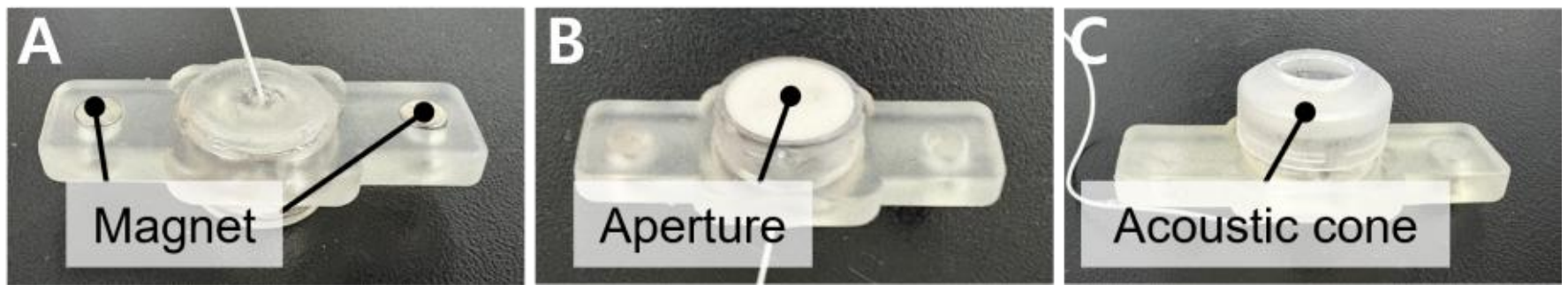

Figure S7: Custom ultrasound transducer. (**A**) Embedded magnet pair enables repeatable and precise transducer placement. (**B**) Transmit aperture. (**C**) Screw-cap acoustic coupling cone for secure mounting.

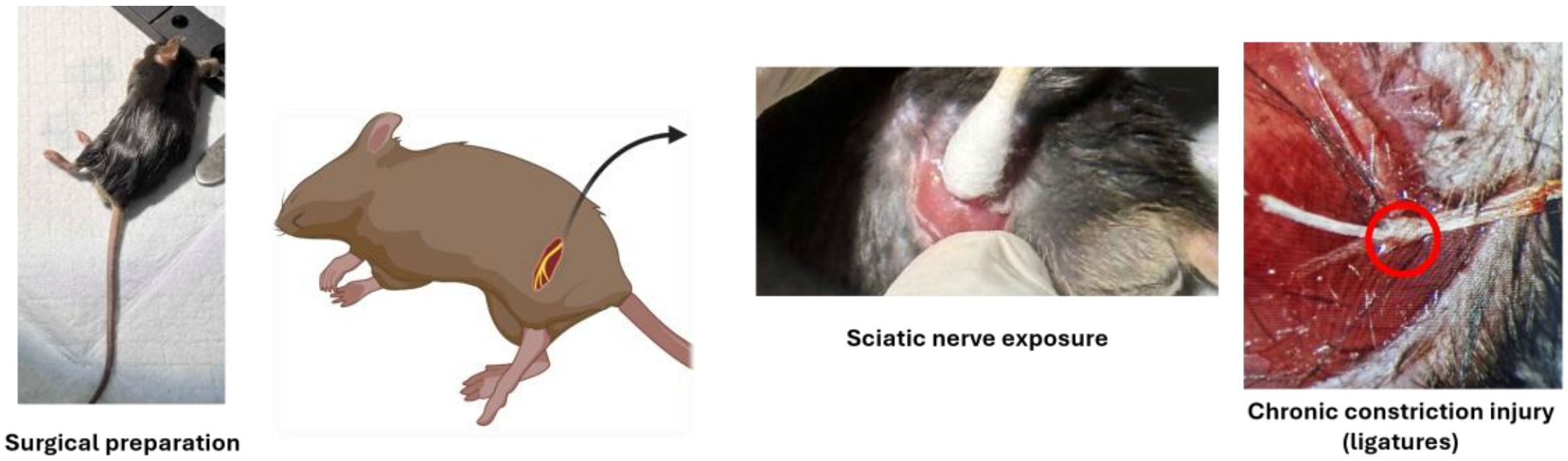

*Figure S8:* ***Representative images illustrating the CCI procedure of the sciatic nerve in mice. From left to right***. *An anesthetized mouse positioned for surgery; schematic illustration of the sciatic nerve exposure site at the mid-thigh level; intraoperative exposure of the sciatic nerve following muscle separation; and loose ligation of the sciatic nerve using 6-0 silk sutures (circled), which induces chronic neuropathic pain. Sham-operated animals underwent identical nerve exposure without ligation.*

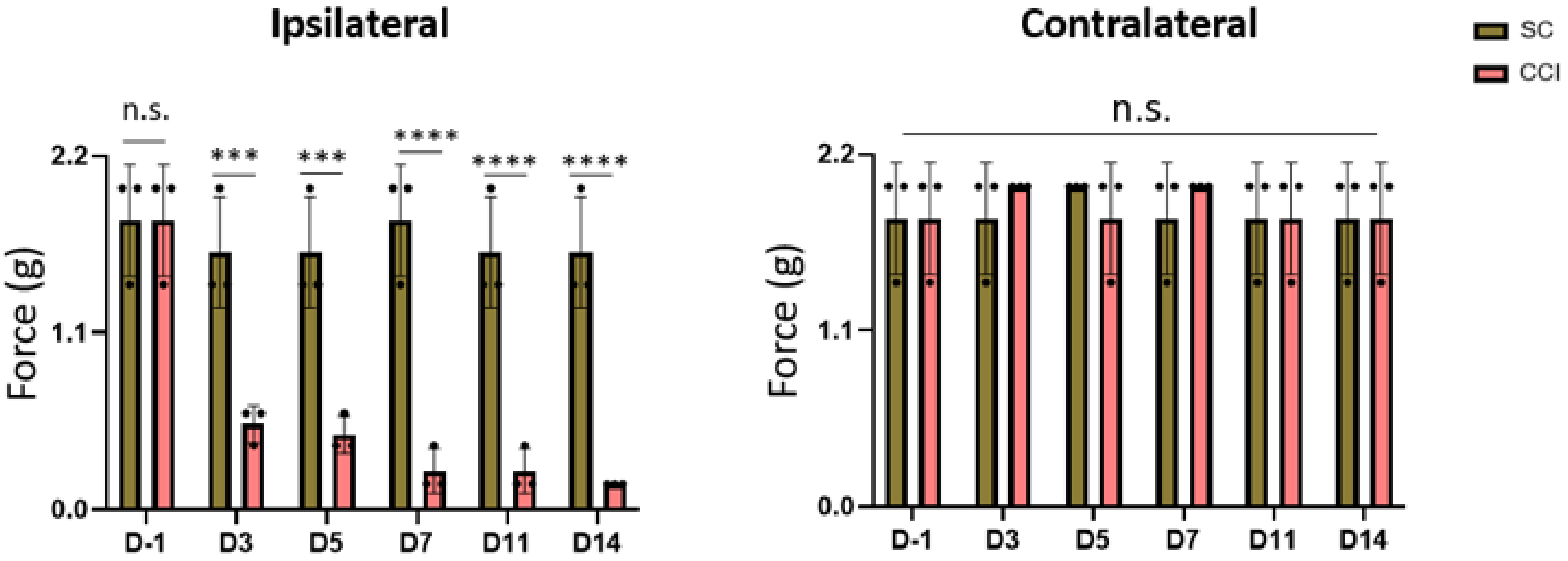

*Figure S9:* ***Progressive development of mechanical allodynia following CCI. Mechanical withdrawal thresholds were assessed using the von Frey test at baseline (day -1) and on days 3, 5, 7, 11, and 14 following CCI surgery.*** *No significant difference was observed between SC and CCI groups at baseline. From day 3 onward, CCI mice exhibited a significant reduction in withdrawal thresholds compared with SC mice, indicating the development and persistence of mechanical allodynia. Dots represent individual animals ($n = 3$ per group). Statistical significance was assessed using two-way ANOVA followed by Sidak's multiple-comparisons test. (***$P < 0.001$, ****$P < 0.0001$; n.s., not significant).*

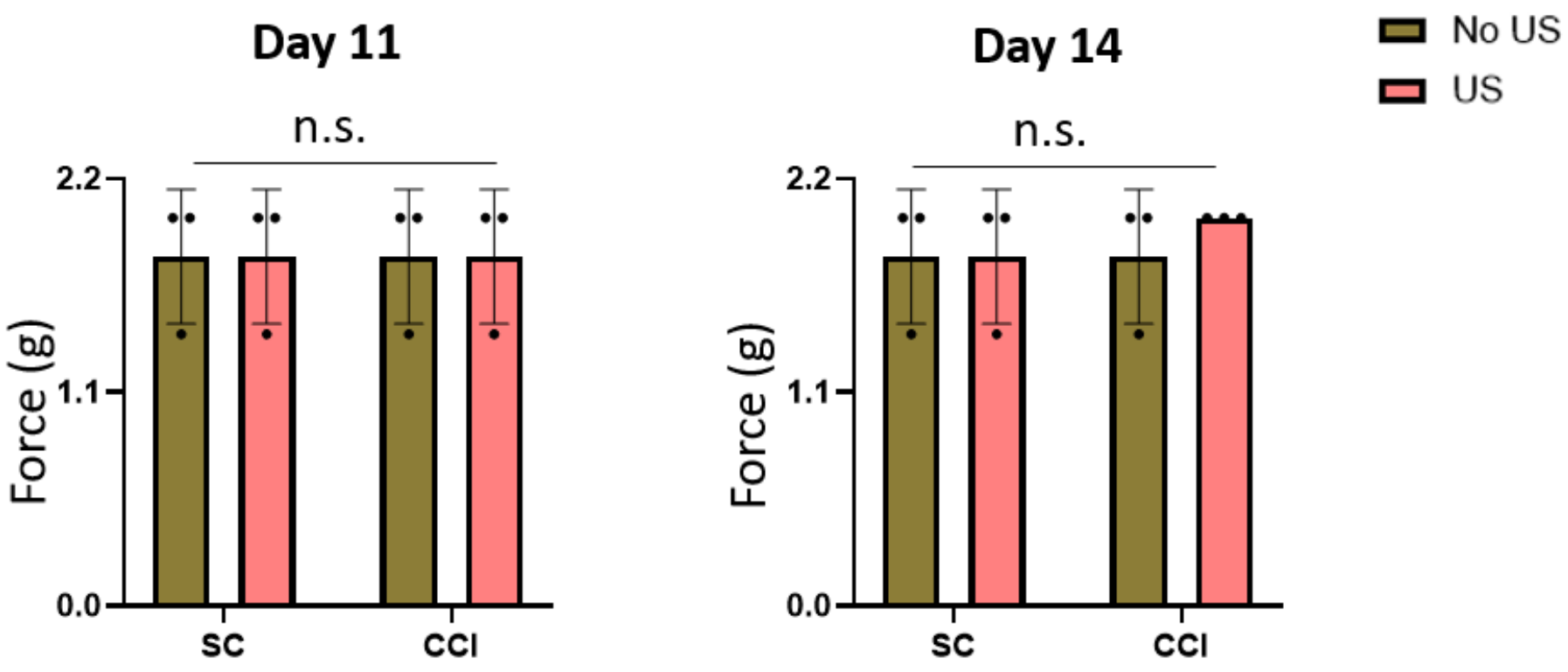

*Figure S10:* ***Contralateral mechanical withdrawal thresholds following ultrasound stimulation.*** *Mechanical withdrawal thresholds were assessed using the von Frey test on the contralateral (non-injured) paw at Day 11 and Day 14. No significant differences were observed among SC, SC + US, CCI, and CCI + US groups at either time point. Data are presented as n=3 in each group, n.s., not significant, two-way ANOVA followed by Sidak's multiple-comparisons test, dots represent individual animals.*

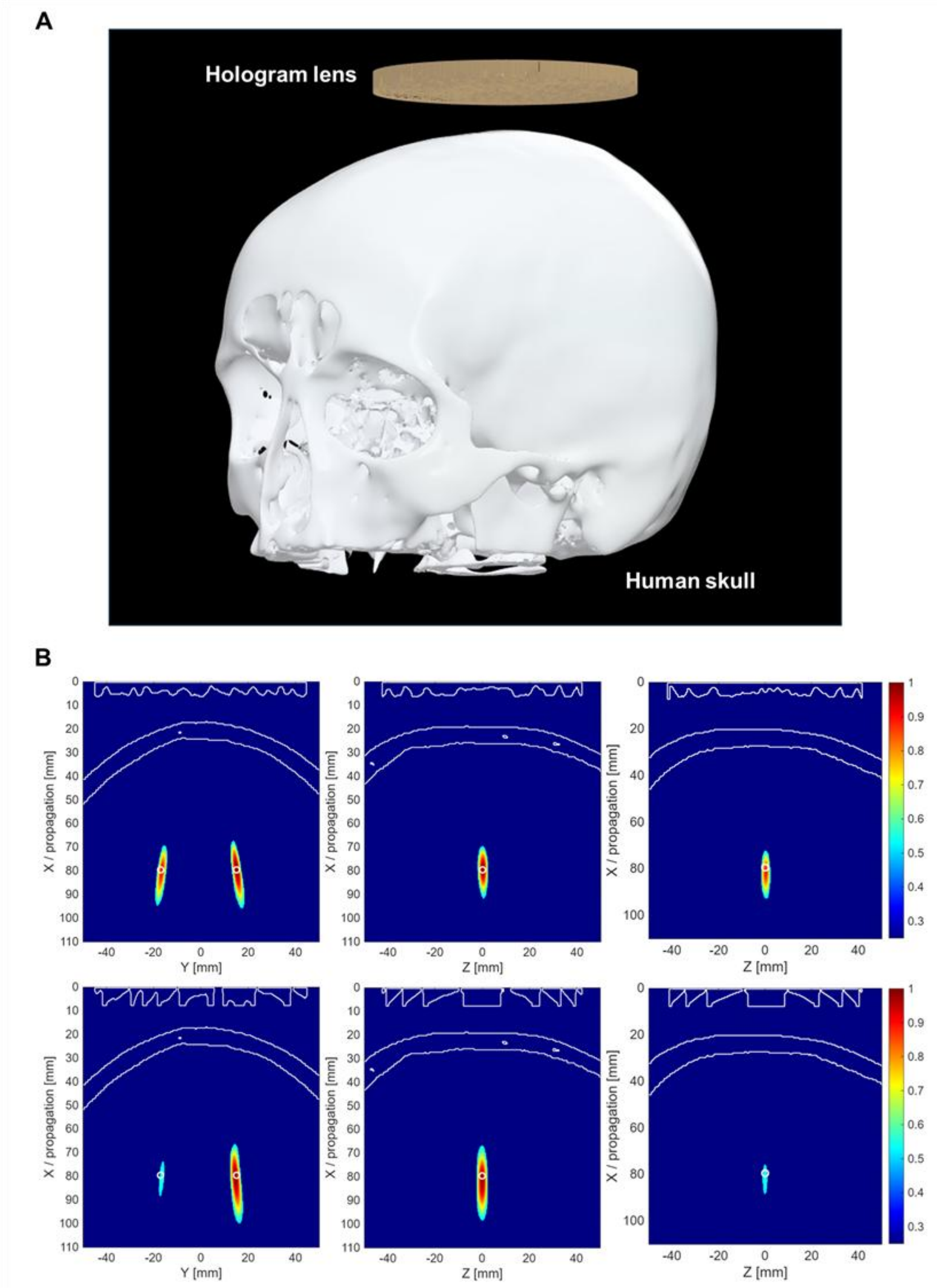

*Figure S11:* ***Human-skull in silico evaluation of the proposed TOAH framework.*** *(**A**) Three-dimensional rendering of the human skull model and the corresponding acoustic hologram lens used for transcranial acoustic field reconstruction. (**B**) Simulated acoustic pressure fields reconstructed through human skull A using TOAH (top row) and TR (bottom row) for the dual-foci configuration at 500 kHz. Cross-sectional views are shown in the XY plane and in two XZ planes corresponding to the two focal targets. White contours delineate the hologram lens and skull boundaries, and white circles indicate the intended focal targets.*

**Table S1: Acoustic properties of the 3D printing materials comprising the lenses.**

| | Sound speed $[m/s]$ | Density $[kg/m^3]$ | Acoustic attenuation coefficient $[dB/(MHz^y \cdot cm)]$ |
|---|---|---|---|
| Form Clear | 2591 | 1178 | 2.922 (y=1.044) |
| VeroClear | 2473 | 1181 | 3.696 (y=0.9958) |
| Agilus30 | 2035 | 1128 | 9.109 (y=1.017) |

**Table S2: Stereotaxic coordinates of the target regions.**

| Region | Anterior–Posterior (AP) | Medial–Lateral (ML) | Dorsal–Ventral (DV) |
|---|---|---|---|
| PAG | -3.200 mm | 0.000 mm | 2.875 mm |
| VP | -1.700 mm | ±1.700 mm | -3.200 mm |

**Table S3: Reported acoustic properties of 3D-printed lens materials tested in this study.**

| Case No. | Reference | Sound speed $[m/s]$ | Density $[kg/m^3]$ | Acoustic impedance $[MRayl]$ |
|---|---|---|---|---|
| 1 | (57,58) | 2424 | 1100 | 2.67 |
| 2 | (59) | 2440 | 1162 | 2.84 |
| 3 | (28) | 2591 | 1178 | 3.05 |
| 4 | (18) | 2700 | 1180 | 3.19 |

**Table S4: Linear mixed-effects model results for the comparison between TOAH and TR in human-skull simulations. Reported coefficients represent the adjusted difference between TOAH and TR (TOAH - TR); positive and negative coefficients indicate higher and lower values for TOAH, respectively.**

| Metric | TOAH vs. TR effect, [95% CI] | P-value | Interpretation |
|---|---|---|---|
| **Cross-domain field accuracy** | +8.88 dB [7.20, 10.55] | **<0.001** | TOAH significantly improved fabrication-domain fidelity relative to TR. |
| **Axial FWHM** | -1.168 mm [-2.33, -0.01] | **0.048** | TOAH reduced axial focal extent overall; however, the pattern of focal compactness differed across frequency conditions. |
| **Lateral FWHM** | -0.06 mm [-0.31, 0.19] | **0.189** | No significant difference in lateral focal width was observed between TOAH and TR. |
| **Focal volume** | -58.58 $mm^3$ [-96.85, -20.30] | **0.004** | TOAH reduced focal volume overall; however, the effect on focal compactness was frequency-dependent. |
| **Focal intensity uniformity** | +0.170 [0.087, 0.252] | **<0.001** | TOAH significantly improved the balance of multiple focal acoustic energy delivery across multiple focal targets. |